\documentclass[10pt]{wlscirep}

\title{Custody Protocols Using Bitcoin Vaults}

\author[1]{Jacob Swambo}
\author[2]{Spencer Hommel}
\author[ ]{Bob McElrath}
\author[ ]{Bryan Bishop}

\affil[1]{King's College London, Department of Informatics}
\affil[2]{Fidelity Center for Applied Technology}

\usepackage{blindtext,graphicx}
\usepackage[absolute]{textpos}
\usepackage[n,
advantage,
operators,
sets,
adversary,
landau,
probability,
notions,
logic, ff, mm, primitives, events, complexity, asymptotics, keys,
]{cryptocode}
\usepackage{array}
\usepackage{algorithm}
\usepackage{algorithmic}
\usepackage{attrib}
\usepackage{mathtools}
\usepackage{hyperref}

\begin{abstract}

A bitcoin \textit{covenant} is a mechanism to enforce conditions on future bitcoin transactions. A bitcoin \textit{vault} is a specific type of covenant transaction that enforces a time-lock on the transfer of control of funds to a hot wallet, but enables an immediate transfer of funds into a deep cold recovery wallet. This paper demonstrates how to integrate a bitcoin vault into a custody protocol and demonstrates the security properties of that protocol. The vault is implemented using pre-signed transactions with secure key deletion (as proposed in \cite{Swambo2020cov}). It is shown that vault-custody protocols enable the wallet owner to specify their desired balance for an inherent trade-off between the security of and accessibility of bitcoin holdings by adjusting the length of time-locks used. It is also demonstrated that wallet owners have increased control of risk-management by compartmentalizing funds across numerous vault transactions. While it isn't realistic to completely prevent theft, the most likely theft scenarios (compromising the hot wallet) have severely limited profitability for an attacker, deterring attempts at theft from the beginning. The proposed architecture was designed to offer defence-in-depth through redundancy and fault-tolerant functionality as well as countermeasures for class breaks through diversity across hardware and software layers. Finally, the architecture employs a detection (a watchtower) and response system that enables fail-safe recovery from attempted or partial thefts through a second type of covenant transaction, a push-to-recovery-wallet transaction.

\end{abstract}

\begin{document}
\begin{textblock}{5}(17,1)
\noindent \today
\end{textblock}

\flushbottom
\maketitle

\thispagestyle{empty}

\section{Introduction}

A custody protocol enables the protection of assets and their functionality and is comprised of software and documentation that supports enacting a set of processes including set-up, deposit, withdrawal, balance-checking, and recovery from failure. The foundations for the custody of bitcoin are key-management (both public and private keys) and an appropriate privacy policy. The robust design, analysis and implementation of custody protocols for bitcoin is a crucial aspect of its capacity to function well as censorship-resistant money since if the set of available custody protocols isn't adequate then users will outsource the protection of their assets. While the space of possible custody protocols is large, in practice only a few are used and most rely on techniques such as multi-signatures and secret sharing \cite{Shamir1979}. Moreover, the space of custody protocols is continually evolving as bitcoin is developed further, with new Script capabilities, new signature schemes, and new transaction validation semantics being proposed as improvements. 

The focus of this paper is on a practical custody protocol based on a promising tool which has yet to be deployed in the industry, namely a \textit{vault} \cite{moeser2016bitcoin}. A vault is a specific type of bitcoin \textit{covenant} that is implemented using pre-signed transactions \cite{P2TST} with secure key deletion, as proposed in \cite{Swambo2020cov}. A protocol such as this offers alternative security properties compared to those based primarily on key-management techniques. In common custody protocols such as multi-signature schemes, keys enable signatures over arbitrary transaction messages, and thus the attack surface when keys are stolen is broad. In this protocol, the idea is to carefully manage pre-signed transactions where the signing keys are deleted. Theft of these pre-signed transactions enable a limited attack surface. Thus funds controlled by pre-signed transactions are less sensitive than those controlled by private keys. The protocol proposed herein uses multiple hardware modules to manage keys and pre-signed transactions.

A bitcoin vault is a tool to enforce gated access to a portion of funds for a specific active (regularly accessed) wallet, while retaining an ability to immediately push those funds to a deep cold wallet (one for which the devices involved have stringent physical and network limitations to ensure high security). It works by committing the funds in `vaulted' custody to being spent according to one of two pathways; either a (relative \cite{BIP68}) time-locked spending path or an immediate recovery spending path. Once the transaction is mined, the funds will not be spendable by the active wallet until the time-lock expires. During this delay period, the access policy can be immediately transformed with a transaction that satisfies the second spending path, collapsing the possible spending paths from two options to one, where the funds become spendable only from a deep cold (recovery) wallet that is only accessed if a partial compromise of the custody protocol is detected or suspected. A vault, then, is a choke point for the flow of control of funds that restricts the number of vulnerabilities presented to an attacker.

\begin{figure}
    \centering
    \includegraphics[width=140mm]{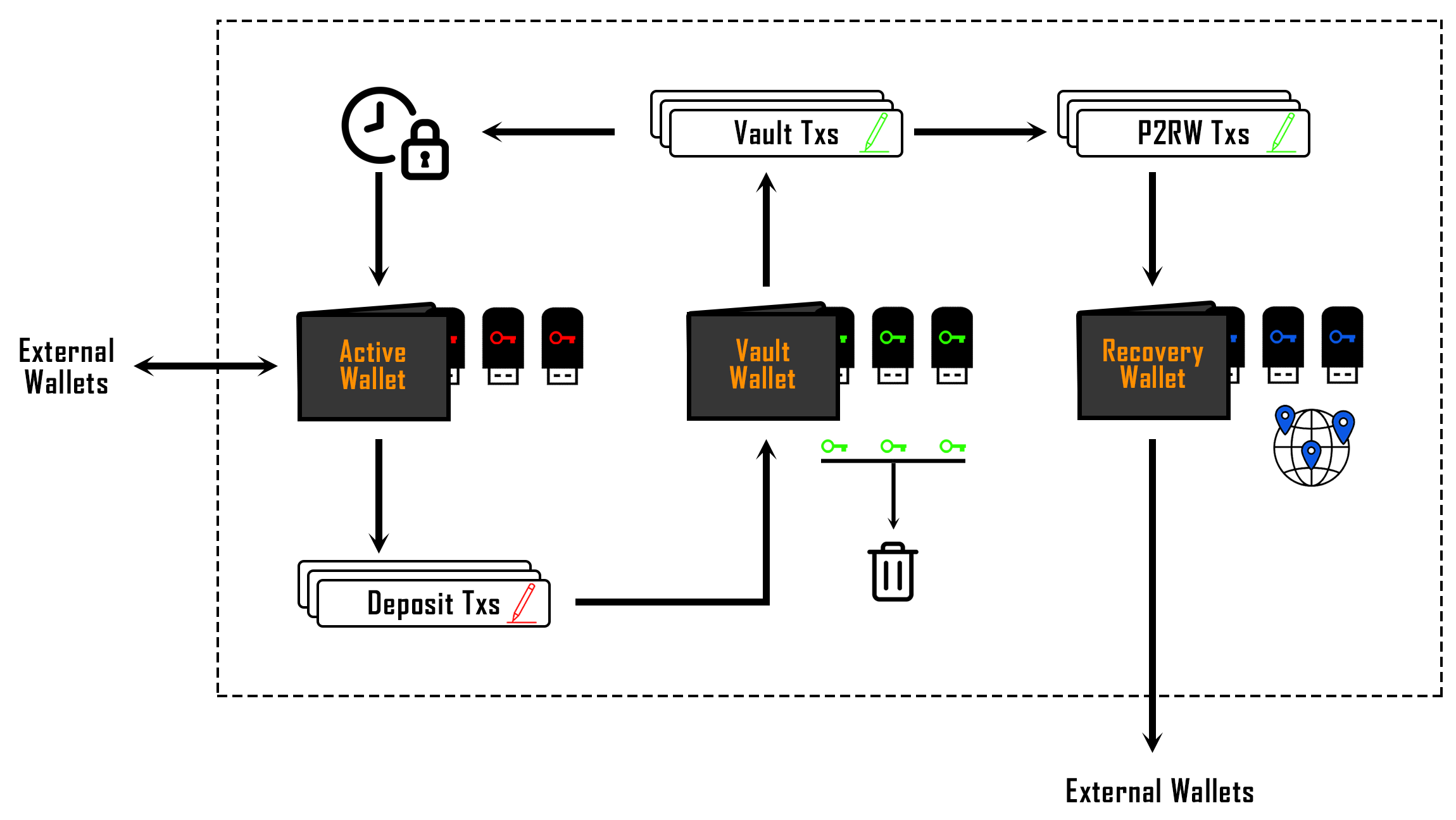} \hspace{2.5cm}
    \caption{Diagram of the access control flow for the vault custody protocol. The active wallet can create arbitrary transactions, and is used as the control point for receiving from and spending to wallets that are external to the custody protocol. It also creates deposit transactions to the vault wallet. The vault wallet uses ephemeral keys to sign vault and push-to-recovery-wallet (P2RW) covenant transactions. A vault transaction has two possible spending paths; it can be accessed by the active wallet only after a time-lock has expired or the P2RW transaction removes the possibility of the active wallet accessing `vaulted' funds and ensures only the recovery wallet can access them. The recovery wallet is a geographically distributed wallet with high physical security that can sign arbitrary recovery transactions moving funds to an external wallet.}
    \label{fig:AccessControlFlow}
\end{figure}

The value of instanciating a relative time-lock in a covenant (rather than a regular time-locked address) is that the time-lock expiration count-down doesn't begin until an attempt to un-vault the funds is made. Since the custodian may monitor the blockchain and peer-to-peer network for evidence of any un-authorized attempt to spend the `vaulted' funds, and if alerted can recover control of the funds by accessing the recovery spending path immediately, attackers are dis-incentivised from attacks that rely solely on the ex-filtration of vault transactions. It is crucial to understand how this affects the strategy space for attackers. An attacker may not gain much from ex-filtrating vault transactions, but may still ex-filtrate private keys from the active wallet. They may simply wait for the wallet owner to un-vault funds, waiting for the time-lock to expire before taking control of those funds. This demonstrates the need for compartmentalization, where funds are distributed through numerous vault transactions that are accessed at a limited rate, and demonstrates the mechanism for limiting profitability from successful attacks that compromise the active wallet. A high-level view of the custody protocol is shown in figure \ref{fig:AccessControlFlow}. It illustrates how the control of funds is managed by transactions between the active, vault and recovery wallets.

{}

The vault custody protocol proposed herein specifies the architectural requirements to use \textit{vault covenants} as a countermeasure for the risk of theft. In general, it is critical that countermeasures operate independently so that the cost of attack is increased, rather than merely shifting the risk to a new set of assumptions under which assets are protected. The requirements for using vault covenants in a custody protocol are an active wallet, a recovery wallet, and a watchtower (whose specific functionality is defined in section \ref{sec:ProtocolFunctions}). This proposal also requires a set of devices (those comprising the vault wallet) to enforce the vault covenant, though this requirement could be alleviated if another mechanism for enforcing covenants in bitcoin becomes available. 

The series of proposals to adjust Bitcoin Script to be suitable for implementing on-chain covenants \cite{moeser2016bitcoin, Covenants2, BIP119} is an ongoing exploration into how to achieve practical bitcoin covenants. This paper adds evidence to the debate and demonstrates that at least some of the desired covenants are practical when implemented using pre-signed transactions with secure key deletion. Modifying Bitcoin's consensus rules is a tremendously difficult process and the mechanism used here does not require any such change. Moreover, if such a change is introduced into bitcoin many of the considerations for a vault-based custody protocol discussed here will still be necessary, as discussed in section \ref{Sec:ScriptCovenants}.

The objective is to design a fault-tolerant and vulnerability-tolerant custody protocol where the loss or compromise of \textit{any single device} at \textit{any time} during any of the processes does not result in a loss of funds. Further, the compromise of the most active set of hardware modules (HMs) should only lead to a limited loss of funds. Under normal operating circumstances, funds controlled by vault covenants are considered the most secure, while funds controlled by private-keys should be transitional only (e.g. small amounts to be spent or external payments to be deposited into vaults). 

The following guiding design principles were inspired from complex systems science \cite{AntiFragileICT} and aim to reduce fragility, promote robustness, and (if possible) promote anti-fragility of the custody protocol. 
    \begin{itemize}
        \item[] \textbf{Modularity:} Components of the protocol (mainly hardware devices) can be lost, compromised, or 
        replaced.
        \item[] \textbf{Weak links:} When components are lost, compromised, or replaced, the overall function of the custody protocol is not broken. A strong link would be a strong dependence between components such that if one brakes the other cannot fulfil its functionality, with too many strong links the global functionality of the protocol would be disrupted by the loss of a single component.
        \item[] \textbf{Redundancy:}  Having multiple components which perform the same functions in the protocol makes the global functionality robust. Single points of failure would otherwise dominate factors that contribute to custodial risk.
        \item[] \textbf{Diversity:} To avoid fragility to malware, software bugs and hardware faults, a diversity of hardware and software should be relied upon within the sets of components that are redundantly performing the same functions. This applies to hardware wallets, software which runs on the hardware wallets, the networked devices and their operating systems. 
        \item[] \textbf{Fault detection:} It should be assumed that faults are inevitable since the environment in which the protocol is instantiated is highly complex. The risks of; malware spreading through networked devices, hardware faults from either supply-chain failures or bit-rot, and malicious human behaviour are ever present. Detection of and response to component failures must happen fast enough to avoid multi-component failures that cause a catastrophic failure of the custody protocol.
\end{itemize}

The remainder of this paper is structured as follows.
Section \ref{sec:Components} introduces the components of the vault custody protocol. Section \ref{sec:Processes} discusses the processes of the protocol, which make use of the different components to enable functionality such as deposits, vaulting funds, un-vaulting funds, recovering from various attacks, and checking the health of components. Section \ref{sec:Threat Model} discusses the threat model for the custody protocol, demonstrating the scenarios in which funds are kept secure, scenarios in which there are limited losses, and scenarios where the protocol catastrophically fails. Section \ref{Sec:ScriptCovenants} gives an overview of the similarities and differences of vault-custody with a script-based covenant mechanism.

\section{Components of the Vault Custody Protocol}
\label{sec:Components}
\subsection{System Architecture}
\label{sec:system-architecture}

\begin{figure}[htp]
  \centering
  \includegraphics[clip,width=140mm]{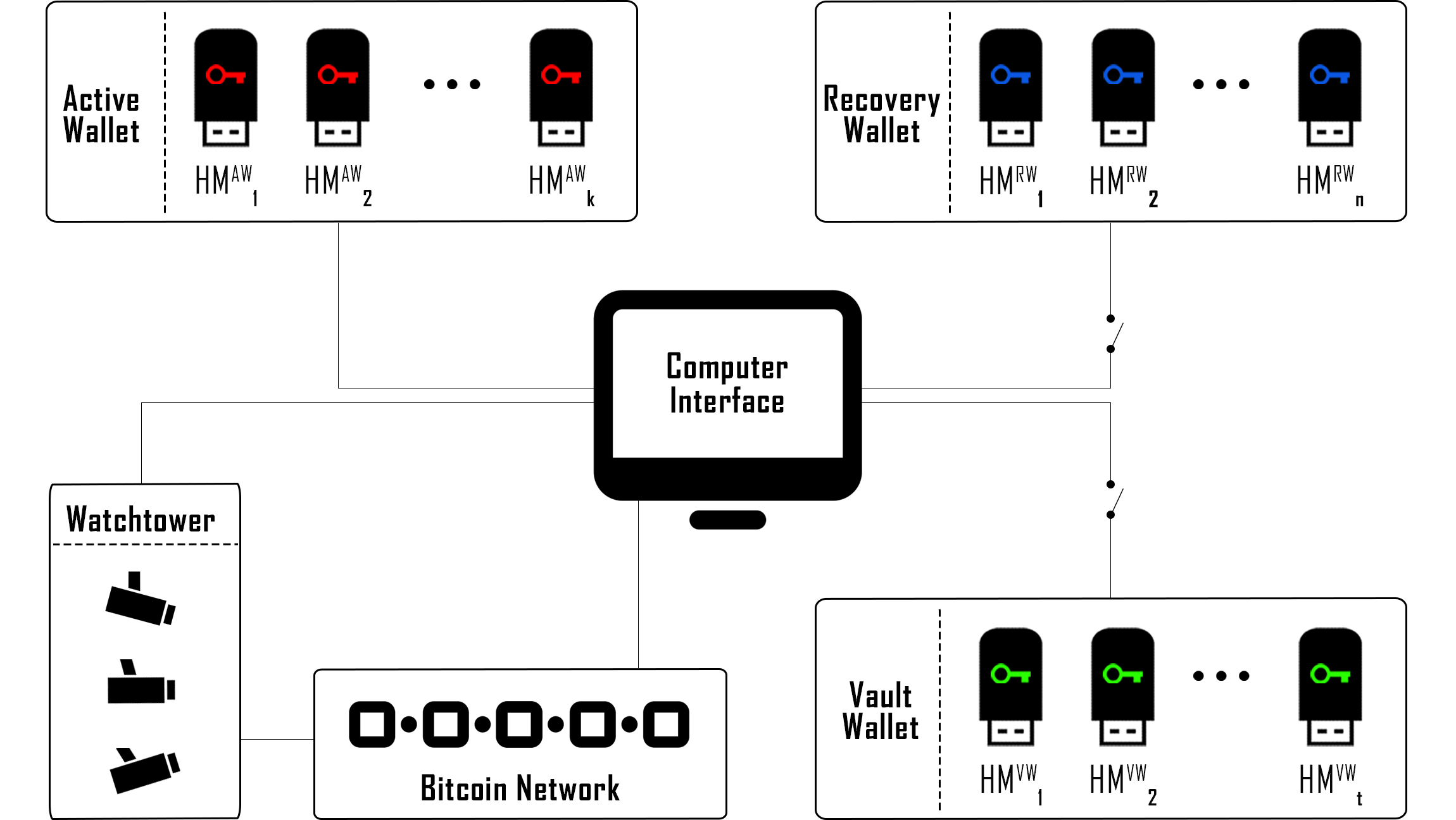}
  \caption{System architecture for the vault custody protocol. The computer interface is a watch-only wallet that interacts with the watchtower, with each wallet type and their hardware modules (HMs), and the bitcoin network. The active wallet is a $j-$of$-k$ multi-signature wallet where the keys are stored on accessible HMs (perhaps even online devices). The recovery wallet is an $m-$of$-n$ multi-signature wallet with eternally quarantined and air-gapped HMs. The vault wallet consists of $t$ HMs which participate in signing vault and push-to-recovery-wallet covenant transactions with ephemeral keys and storing those covenant transactions. The watchtower is a distributed set of full nodes that detect events relevant for the wallet owner by monitoring the bitcoin network and warn of or respond to those events.}
  \label{fig:Architecture}
\end{figure}

Figures \ref{fig:Architecture} shows a system architecture for the vault custody protocol. The system relies on heterogeneous multi-signature wallets that use numerous hardware modules (HMs).
The active wallet is used often and is connected to the computer interface making it a riskier wallet to grant control of funds to than the recovery or vault wallet. The vault wallet is used to endorse and manage \textit{active covenant transactions}\footnote{In keeping with the definitions given in \cite{Swambo2020cov}, once the necessary signing keys have been used to create the signatures are deleted and the deposit transaction is confirmed, the covenant transactions become \textit{active}.} (ACTs). The recovery wallet is accessed as rarely as possible, only in the event of emergencies and device health-checks. 

The choice of hardware wallet devices is critical. Where standard hardware wallet functionality is required, industry leading devices are considered (e.g. Trezor, Ledger). Where additional functionality is required to enable secure key deletion and the storage of ACTs, custom hardware is required since there is nothing commercially available. Note that the software for this custom hardware is under development and is open-source. It is being used to prototype this custody protocol and can be found at the \href{https://github.com/fmr-llc/Vault-mbed}{vault-mbed github repository}.

\subsection{Protocol Functions}
\label{sec:ProtocolFunctions}

The computer interface interacts most often with the active wallet HMs, but also infrequently functions as the interface between all other HMs and the watchtower. It mediates secure communications between distributed wallet HMs which must communicate sensitive information between each other (such as public-keys and partially-signed vault transactions).

The active wallet HMs collaboratively perform the typical distributed key generation and signing functions of a multi-signature wallet. The active wallet functions include receiving external payments\footnote{The intention is to separate receiving payments from depositing funds into a vault because payees cannot be guaranteed to follow any given prescription for payments. It would create a serious flaw in the custody protocol if a payee attempted to re-use an address for which the private keys had been deleted to activate a bitcoin vault.}, spending funds with arbitrary transactions, depositing funds into the vault wallet, and participating in un-vaulting (by adding fees to ACTs that are to be braodcast). This puts a responsibility on the active wallet HMs to maintain any essential redeem scripts for the ACTs as well as funds held in multi-signature addresses. 

There is a dedicated set of HMs to perform the vault wallet functionality; signing two types of covenant transactions for vault construction (whose design is described in \ref{vault-design}), storing them (\ref{sec:AVT Storage}) and securely deleting the signing keys (\ref{sec:secure-key-deletion}) to make them \textit{active}.

The recovery wallet HMs constitute a multi-signature wallet with the highest device security through minimized usage, physical security measures, eternal quarantine and geographical distribution. They should only be accessed if a recovery process has been triggered or if a device health-check is to be performed. Their function is to take control of funds \textit{after} the recovery process has completed; control of funds can be redirected from vault outputs to a recovery-only output with a push-to-recovery-wallet covenant transaction (\ref{vault-design}), and funds controlled by the active multi-signature wallet can have their control transferred to the recovery wallet HMs through a regular transaction. At the end of the recovery process, the recovery wallet HMs become the new active wallet HMs and a new (or recycled) set of HMs become the recovery wallet hardware. 

The watchtower operates several full bitcoin nodes which monitor the blockchain and peer-to-peer network for events relevant to the wallet owner. In particular, the watchtower guards against attempted thefts of funds secured in a vault through unauthorized broadcast of vault transactions. The watchtower requires a resilient design with fault-tolerance since the security of funds may depend largely on the watchtower. A detailed discussion of the watchtower's requirements is given in \ref{sec:watchtower-design}.

\subsection{Protocol Information Architecture}
\label{sec:InformationArchitecture}

There are several types of information to be protected within the vault custody protocol; private signing keys, public addresses, ACTs, redeem scripts, authentication keys, and information required to establish communication channels. Each type of information has different storage requirements given the context of its use and the consequences of losing that information or having it stolen. 

Each HM will protect sets of private keys whose entire lifecycles are contained on that hardware device. At no point will private keys, whose role is to sign bitcoin transactions, be intentionally transferred to the computer interface or other device. The private keys that enable the active wallet and recovery wallet functionality have much longer lifecycles than the ephemeral private keys used to sign vault and push-to-recovery-wallet covenants. The security of the individual private keys on each HM depends mostly on trusted hardware and appropriate procedural controls. By relying on a distributed multi-signature set-up for the active, vault, and recovery wallets, there is a tolerance to faults and thefts of individual HMs that will be explicitly defined in section \ref{subsec:Functions and Tolerances}.

For the active and recovery wallets, sets of multi-signature public addresses should be securely stored. These addresses are composed of public keys that are generated on the set of HMs for the associated wallet type (active or recovery). These addresses must be redundantly stored for fault-tolerance and must be readily accessible for payment and recovery processes. The protocol proposed herein redundantly stores the public addresses for active and recovery wallets on their associated set of HMs. The recovery wallet addresses will also be redundantly stored on the active wallet HMs for ease of access when constructing new P2RW transactions, without compromising the recovery HMs by requiring regular access to them. This relies on secure communication channels between the HMs of the wallets (intermediated by the computer interface) as described in section \ref{sec:SecureCommunication}.  

The vault wallet uses ephemeral private keys that only exist during the vaulting process (described in section \ref{sec:Vaulting}) on the vault wallet HMs. The associated public keys are also generated during the vaulting process, and passed to the active wallet to form multi-signature deposit addresses which vault transactions will spend from. Covenant transactions, once active, are to be distributively stored on some chosen number of vault wallet HMs. Covenant transactions are generated in pairs where each vault transaction has an associated P2RW transaction. Through the vaulting process (from partially-signed to active) the vault transactions should never be kept on a networked device such as the computer interface. The P2RW transactions must be quickly accessible to handle detected intrusions and will thus be stored by the computer interface, or with each of the watchtower nodes.

Each ACT will have an associated redeem (witness) script which should be redundantly stored on the active wallet HMs. These redeem scripts must be accessible. These redeem scripts are needed by the active  wallet HMs when claiming the un-vaulted funds through either the time-locked spending path or with the recovery spending path.

The authentication keys for secure communication across multiple channels between the wallet owner and the watchtower should be stored by them both. The authentication information for secure communication between the set of HMs for each wallet type should be stored on each of the relevant HMs. 

\subsection{Key Structure}

\label{sec:KeyStructure}

Each HM in the active and recovery wallet will generate its own key-tree for a hierarchical deterministic wallet as per BIP-0032 \cite{BIP32}. The vault wallet will not be a hierarchical wallet since the private-keys are supposed to be ephemeral and must not be able to be re-derived to ensure the security of the covenant. To further specify and standardize deterministic wallet structure (to improve interoperability of different wallet softwares) BIP-0043 \cite{BIP43} was drafted, and suggests the use of a \textit{purpose} field in the key-derivation path of the tree structure. Given the specific wallet structure required for the vault custody protocol proposed herein, a new purpose field is defined to be \textit{vault custody}. In keeping with the intended format for further extending the key-derivation path for flexible and standardized wallet structures, a similar approach is taken here as to that of BIP-0044 \cite{bip44}. There are three wallet types used in this protocol including the active, vault, and recovery wallets. Of these, the active and recovery wallets are multi-signature and it would be most convenient if public-keys and multi-signature addresses could be derived without necessarily having access to each individual HM. For this, non-hardened public keys are used\footnote{A hardened derivation of public keys requires access to an associated private key from the key-tree \cite{BIP32}}. Thus, following the vault custody purpose field a \textit{wallet type} field is defined which can take either of two values; \textit{active} or \textit{recovery}. The non-hardened public keys will need to be securely stored on the HM devices in such a way that the public keys cannot be maliciously mutated nor leaked from the device. The key-tree derivation path takes the following form (using the shorthand notation from \cite{BIP32}):

\hspace{4cm}{\tt m/vault custody/wallet type/}

\subsection{Secure Communication}
\label{sec:SecureCommunication}

Private communications between HMs are necessary to maintain privacy and to guard against a corrupted computer interface. Since the computer interface is significantly more vulnerable to network attacks, its access to the public-key and derivative information should be limited as much as is possible. The use of non-hardened public-key derivation for the active and recovery wallets means that a privacy loss of one key affects all derivative keys. If this information is leaked, an attacker will be able to infer operational information of the custody protocol being used (through blockchain analysis and inferences based on vault scripts in use). 

A suitable cypher suite  for secure communication between devices would be TLS-ECDHE-PSK-WITH-AES-256-CBC-SHA384 (\href{https://tools.ietf.org/html/rfc5489}{RFC 5489}). Given that HMs will have support for elliptic-curve cryptography over the curve Secp256k1 using Elliptic Curve Ephemeral Diffie Hellman (ECDHE) is an obvious choice for the key agreement protocol. Relying on pre-shared keys (PSK) for the communication protocol alleviates any reliance on a certificate authority\footnote{One could also use TLS with Secure Remote Password to avoid certificate authorities and enable mutual authentication \cite{SRPTLS}. Alternatively, one could act as their own private certificate authority and self-sign certificates for each device \cite{OpenSSLDocs}.}. The pre-shared key can be input by a human operator during the set-up process (see \ref{sec:SetUp}). For strong encryption, AES-CBC with a key size of 256 bits can be used and SHA-384 is used for message authentication. 

\subsection{Vault Design}
\label{vault-design}

A class of bitcoin covenants were shown to be implementable without modification to bitcoin's consensus rules in \cite{Swambo2020cov}. The way it is done is to construct a transaction which encapsulates the specific spending conditions permitted by the covenant, sign the transaction with ephemeral keys (used only for this task), securely store those pre-signed transactions, and then delete the signing keys. The vault design presented herein relies on two types of covenant transaction; a vault covenant and a push-to-recovery-wallet (P2RW) covenant. A vault covenant is a specific sub-class of covenants in which the locking script for the output has a specific function: the funds can either be spent to a pre-defined address after the expiration of a time-lock, or the funds can be immediately redirected to a recovery address. P2RW transactions are simple covenants which commit funds to be transferred to a specific target address, in this case, a multi-signature address controlled by the recovery wallet. This push-to-recovery-wallet covenant transaction is used to transform the access control of funds controlled by a vault covenant from having two possible spending paths (one time-locked, and one immediate) to having a single immediate spending path. The benefit of this approach is that the recovery wallet signing keys don't need to be accessed in order to recover funds. This requires a transitional multi-signature address (consisting of public keys generated by the vault wallet) as the target of the vault transaction's recovery spending path.

The design space for vaults made in this way is still broad. To simplify, the version, lock-time, and outputs (including the vault output and a change output) are specified. The version field of the transaction is set to 2, meaning that the relative lock-time feature is available \cite{BIP68}. The transaction-level lock-time field is specified with a number that represents a time that has already passed so that the transaction-level lock-time is not active. The first input is set and can be thought of as the deposit to the vault. It is critical that vault transactions are non-malleable since the push-to-recovery-wallet transactions depend on their static transactions IDs. The vault transaction should thus be a Segregated Witness transaction whose funding input is signed with {\tt SIGHASH\_ALL}. The input's signature commits to the version, lock-time and all outputs and inputs. 

A vault transaction output script has the following form, and is used to construct a pay-to-script-hash (P2SH) or pay-to-witness-script-hash (P2WSH) address: 

\begin{verbatim}
          OP_IF
              T OP_CHECKSEQUENCEVERIFY OP_DROP 
              j <active wallet pubkey 1> ... <active wallet pubkey k> k 
              OP_CHECKMULTISIG
          OP_ELSE
              p <recovery path pubkey 1> ... <recovery path pubkey t> t 
              OP_CHECKMULTISIG
          OP_ENDIF
\end{verbatim}{}

This vault script has a (relative\footnote{Given a transaction with version 2, as specified in BIP68 \cite{BIP68}.}) time-lock of length T. The vaulted funds can only be spent with a sufficient threshold of signatures created with the private keys associated with the active wallet public keys ($j-$of$-k$) or the recovery path public keys ($p-$of$-t$). To access the IF execution branch of the script one would construct an input that points to the vault output and provide signatures and an unlocking script (scriptSig) of the form

\begin{verbatim}
          <active wallet signature 1> ... <active wallet signature j>
          OP_1 <redeem script>
\end{verbatim}{}
where {\tt OP\_1} accesses the IF branch and the redeem script is the locking script of the vault output. To access the ELSE execution branch the scriptSig uses {\tt OP\_0} and has the form: 
\begin{verbatim}
          <recovery path signature 1> ... <recovery path signature p>
          OP_0 <redeem script>
\end{verbatim}{}

This ELSE branch is used for the push-to-recovery-wallet covenant transaction with an input that is signed with $p$ \textit{ephemeral} recovery path private-keys. The P2RW transaction creates a new output with a locking script similar to the following:

\begin{verbatim}
          m <recovery wallet pubkey 1> ... <recovery wallet pubkey n> n
          OP_CHECKMULTISIG
\end{verbatim}{}

This shows an $m-$of$-n$ multi-signature locking script for the recovery wallet. The P2RW transaction can be pre-signed with the SIGHASH flag {\tt ALL|ANYONECANPAY} such that new inputs can be added to dynamically allocate fees. This requires a fee-paying wallet that has a well structured set of unspent outputs that would require no change. One might consider using {\tt SINGLE|ANYONECANPAY} to enable adding additional change outputs to the transaction, for better flexibility with the fee-paying wallet, but this makes transaction pinning attacks possible \cite{Pinning}.

\subsection{ACT Storage}
\label{sec:AVT Storage}

Once a covenant transaction is activated (the signing key deleted and the deposit confirmed), broadcasting the covenant transaction is the only way to access its funds. There are two desirable properties for the secure storage of ACTs; (1) to be resistant to theft and (2) to be resistant to accidental loss. These two properties introduce a trade-off in the design space because adding redundancy to the storage of ACTs to improve `resistance to loss' introduces a broader attack surface, which diminishes `resistance to theft'. However, priority should be given to resistance to loss since this failure mode is final. Theft of transactions enables an early broadcast, but does not immediately result in loss of funds. 

While funds are inherently safe from theft when stored as vault transactions since the recovery spending path is immediately accessible, it is undesirable to trigger the recovery process because it involves re-creating (at least) part of the protocol set-up, with new vault deposits, etc., as discussed in section \ref{sec:SetUp}. Observe that accidental loss of the active vault transactions (AVTs), for example through HM malfunction, would result in total loss of funds. 

The obvious solution to ACT storage is to create multiple redundant backups, distribute them across trusted storage devices, and depend on the watchtower to protect against theft of vault transactions. Tuning the number of copies requires satisfying the risk tolerance of the custodian. The vault wallet HMs are suitable for storing the vault transactions (and back-ups of the P2RW transactions). However, the P2RW transactions should be easily accessible in case an intrusion is detected or suspected. These may be stored by the computer interface or the watchtower nodes to enable triggering the recovery process.

While it would be possible to use a verifiable secret sharing \cite{VSS} scheme to distributively store the ACTs, increasing the resistance to theft, this would add significant complexity to the protocol and its analysis and so is left for future research.

\subsection{Secure Key Deletion}
\label{sec:secure-key-deletion}

Key deletion is a necessary part of activating the covenant transactions used in the custody protocol proposed herein. Without \textit{secure} key deletion, the key may be recovered by an attacker and be used to sign transactions that send funds to addresses not specified in the covenant, thus breaking the covenant. Secure deletion has received significant attention in cryptography literature, for reference see the book \cite{SecureDataDeletionBook}.

Realistically, secure key deletion on a single device is not reliable. HMs can be compromised by side-channel attacks, supply-chain attacks and malware to name a few. As discussed in \cite{Swambo2020cov} one can improve the security of key deletion by having a vault transaction's deposit be locked with a multi-signature Script, where the private keys required are each stored on different hardware devices which have to collaboratively sign the vault transaction to verify it. Under an $n$-of-$n$ multi-signature scheme only one of the hardware devices needs to successfully perform the key deletion (without it being recoverable) to guarantee the vault covenant. This approach (or a similar threshold signature approach) is essential for ensuring there is no single point of compromise for the secure key deletion. However, it adds a layer of complexity to the protocol since the separate vault wallet HMs need to communicate in order to receive and sign a partially-signed covenant transaction, and unfortunately this increases the lifecycle of any private keys used to sign them.  

To increase the cost of attacks  that use malware or compromise the supply-chain one would ideally use multiple operating systems for the devices which will be deleting keys and securing ACTs. This diversity acts as an additional layer of security. Moreover, diversity could be introduced at the hardware and software layers too; relying on devices that perform the same functions but on different substrates makes generic malware attacks more difficult.

For the purpose of prototyping, the approach taken herein is to use $p$-of-$t$ multi-signature addresses for both types of covenant transactions. Key deletion is secure if at least $p$-of-$t$ of the signing devices resists compromise and successfully delete their signing key. 

\subsection{Watchtower Design}
\label{sec:watchtower-design}

The watchtower forms a critical part of the security model for the vault custody protocol. Since control of vaulted funds is time-locked, and is in a sense outsourced to the consensus protocol, it is inherently an online security model and thus requires an online watchtower. This discussion will give an overview of four different types of watchtower that could be integrated into a custody protocol that uses vaults, but a full specification is left for future work. There are two dimensions of design worth considering. First, whether or not the watchtower is given the responsibility of triggering the recovery process (by broadcasting P2RW covenant transactions). If not, it would be a simple watchtower that monitors the bitcoin network and notifies the wallet owner when an attempt to un-vault funds is made. The second dimension is whether or not the watchtower is outsourced. In each case (non-outsourced notifying, non-outsourced responding, outsourced notifying, outsourced responding), the precise objectives of the watchtower differ, as well as the security and privacy properties offered to the custody protocol. 

\subsubsection{Watchtower Fundamental Design}

There are aspects of the watchtower which are common to each design. A watchtower must operate as a full bitcoin node to enable independent validation of transactions (in particular, the validity of vault and P2RW transactions). A single full node would be vulnerable to an array of modes of compromise and failure (e.g. eclipse attacks, power outages, etc.). Thus a robust watchtower should consist of multiple nodes; this reduces the risk of downtime since only one of the nodes needs to operate successfully to maintain the security of the custody protocol. The risk of other generic faults may be reduced by running nodes with different operating systems and hardware. In addition, each node must maintain connectivity to a large and diverse set of peers in the bitcoin network (a patch may be required to enable more aggressive peer discovery than is standard in bitcoin core). Finally, multiple independent communication channels must be established with the wallet owner. The watchtower isn't useful if it can't communicate with the wallet owner securely. The use of multiple channels adds redundancy to the communication and increases an attacker's cost to compromise the communication.  

\subsubsection{Notification Watchtower}

The objective of a notification watchtower is to continuously scan the peer-to-peer network and blockchain for the appearance of any of the set of vault transactions for which it has been tasked with monitoring, and to notify the wallet owner if any valid vault transactions are detected. For this, the watchtower must have identifying information such as the {\tt txid} of the vault transactions. The simplest watchtower would notify the wallet owner of \textit{any} attempt to un-vault funds, even those broadcast by the wallet owner. The wallet owner would then discern whether or not an attempt at theft was made. 

In addition to detecting any broadcast of known AVTs, the watchtower may scan the network for activity related to any unspent outputs of relevance to the wallet owner. In particular, due to the possibility of `re-packaging' inputs and outputs from covenant transactions into different transactions, the watchtower would function better if programmed to watch for attempts to spend unspent outputs from vault deposit transactions. Other notification types might include warning a user that the rate of broadcast of AVTs is too high, and that this puts a large portion of funds at risk.

\subsubsection{Responder Watchtower}

Upon detection of an unauthorised but valid attempt to un-vault funds, the P2RW should be broadcast before the vault time-lock expires. These P2RW transactions can be given to the watchtower such that it has the responsibility of triggering the recovery process. A responder watchtower has the objectives of;
\begin{itemize}
    \item[(a)] verifying the authenticity of a notification from the wallet owner that a vault transaction will be broadcast for un-vaulting,
    \item[(b)]detecting events relevant for the wallet owner (as with the notification watchtower above), 
    \item[(c)] securely storing the P2RW transactions, and
    \item[(d)] triggering a recovery process when necessary by broadcasting P2RW transactions.
\end{itemize}{}

The authenticated un-vault messages depend on secure communication between the wallet owner and the watchtower, established during the set-up process (section \ref{sec:SetUp}). In addition to storing the set of {\tt txids} for the vault transactions, a responder watchtower must also securely store P2RW transactions. These transactions should be stored redundantly (by the watchtower and perhaps elsewhere by the wallet owner) to resist accidental loss. A compromised set of P2RW transactions enables an attacker perform a form of denial-of-service (DoS) attack by triggering the recovery process, causing some grief for the wallet owner (see Section \ref{sec:Threat Model}). However, assuming the recovery wallet remains secure, the attacker will not be able to gain control over funds by only compromising a watchtower node. A responder watchtower should warn the wallet owner if it doesn't have access to the necessary P2RW transactions.

\subsubsection{Outsourcing a Watchtower}

The watchtower may require some professional expertise and infrastructure to operate. If the watchtower functionality can be securely outsourced then the use of the vault custody protocol would be more accessible to users with less expertise and equipment. So it is worth considering the consequences and security properties for outsourcing a watchtower. Two major concerns for relying on an outsourced watchtower relate to corruptibility and privacy.

An outsourced watchtower must be informed of some identifying information for the vault transactions that control the funds it is monitoring. Initially, knowing the {\tt txid} does not compromise the privacy of the wallet owner and their account information. However, once the transaction is broadcast, the watchtower will gain information on the input of the vault transaction. Once the funds controlled by it have been spent from the watchtower will have further information on the output Script of the vault transaction. A responder watchtower will also have information contained in the P2RW transactions. If the operator of the watchtower has no identifying information about their client then this may not be a problem. 
However, if a watchtower operator collects data from a \textit{known} client (e.g. identified through email, IP address, or phone number), then over a reasonable period of time they could begin to infer some information that should remain private for both the security of the custody solution and the fiscal privacy of the wallet owner. 

Watchtower service providers will offer different payment models. Clients could be charged for each vault transaction it monitors, or charged for a subscription, or perhaps there is some on-chain payment mechanism integrated with the P2RW transaction. It is unlikely that clients would be charged for each attempted theft, since the presence of a watchtower deters attackers from broadcasting stolen vault transactions. An incentive to operate honestly and correctly could be based on its reputation and the long-term benefits of its profits. The service provider could however be bribed, attacked, or fail in some other arbitrary way. Compared to a non-outsourced watchtower, the main difference here are threats that mis-align the incentives of the service provider with those of the wallet owner (e.g. corrupting bribes). 

As with a non-outsourced watchtower, there is still a requirement for running multiple nodes. An outsourced watchtower functionality would benefit from hiring multiple independent watchtowers (who aren't aware of each other) to reduce the risks of malicious behaviour. Again, only one watchtower needs to correctly function to maintain protection from theft. On the other hand, hiring more watchtowers reduces the privacy of the wallet owner. Some obfuscating information could be passed to watchtowers to regain some privacy, such as irrelevant vault and P2RW transactions, though this would require careful and thorough design and does not handle the fact that information is eventually revealed on the public ledger.

\subsection{Mitigation for Corrupted Computer Interface}
\label{sec:OOBMutAuth}

Assuming the computer interface can be corrupted, there are two known attacks on the HMs that the computer interface communicates with that require mitigation; payment attacks and address generation attacks \cite{FormalHardware}. A payment attack is when a corrupted computer interface passes a malicious transaction to the HM for it to sign. An address generation attack is when a corrupted computer interface adds a public address controlled by an adversary to the list of receiving addresses. Both of these attacks are typically mitigated by having a human-check that the HM and computer interface are behaving as intended (i.e. that transactions have not been maliciously modified and that addresses held by the computer interface match those controlled by the HM). The human-check compares data across a visual channel, typically a display on the HM and on the computer interface. The visual channel is a type of out-of-band (OOB) communication channel, one which is independent of the primary in-band channel \cite{Latvala2019} (in this case the communication between HM and computer interface). The OOB channel expands the effort required by an attacker to enact a payment or address generation attack since now both the in-band and OOB channels must be compromised.

Given the heavy reliance on multi-signature security of the protocol proposed herein it is worthwhile considering how to minimize the number of necessary human-checks. Requiring access to numerous devices each time a payment is to be received or sent is cumbersome. However, given secure communications between HMs it can be assumed that there will be consistency across devices for sets of multi-signature addresses that are stored for the active and recovery wallets (e.g. each HM in the active wallet will hold all of the active wallet multi-signature addresses and/ or public keys). Thus to avoid an address generation attack, only one HM from the relevant set of HMs needs to be accessed and checked. 
Similarly, each time a wallet is to engage in a distributed signing protocol of a transaction, provided the partially signed transaction is passed securely (encrypted) between the HMs, then a human-check is only necessary for the final signature which fully verifies a transaction. 

\section{Processes of the Vault Custody Protocol}
\label{sec:Processes}

\subsection{Set-up}
\label{sec:SetUp}

This process describes how each of the devices should be set up and configured, how the relevant initial information should be prepared and secured for receiving external payments, for recovery processes, and for vault deposits.

The computer interface intermediates secure communications among and between each set of HMs. To establish secure communication channels pairwise between each HM, an operator does the following for each channel. Generate a key on one of a pair of HMs. Read this key off the HM screen and input it directly onto the other HM with which to create the channel. Connect both HMs to the computer interface to mediate the TLS-PSK handshake \cite{PSKTLS} and establish secure communication. The goal is to allow the HMs to share and redundantly store the relevant multi-signature addresses without revealing this information to the computer interface\footnote{Notice that the HMs require secure storage functionality to maintain the multi-signature addresses without risk of them being mutated.}. The secure communication will also be relied upon during the vaulting process described in section \ref{sec:Vaulting}.

Each HM will perform its own key-tree generation according to the key-structure specified in \ref{sec:KeyStructure}. The next step is to construct the relevant multi-signature addresses for the active and recovery wallets among the associated set of hardware devices. The relevant HMs will communicate their public-keys and construct a consistent set of multi-signature addresses and store them as specified by the information architecture in section \ref{sec:InformationArchitecture}. This distributed key-generation process is done without ever bringing the private-keys onto a single device (and ideally not to a single location).

As specified by the information architecture of section \ref{sec:InformationArchitecture}, the active wallet HMs will store the multi-signature addresses for the active and recovery wallets. The recovery wallet HMs should also store its addresses to be able to validate the transactions that it will provide signatures for during the recovery processes. The vault wallet HMs should also store the recovery and active wallet addresses so that it can verify the correctness of the covenant transactions it is signing.

Standard physical security procedures for the HMs should be enacted once they have been configured (including the use of safes, armed security, CCTV, \textit{etc.}). The recovery wallet HMs must be held in separate secure locations, and should not be accessed except for emergencies and device health checks. 

The watchtower should be initialized with the necessary public information to be able to authenticate messages from the wallet owner. End-to-end encrypted communication can be set-up through Diffie-Hellman key exchange with each node of the watchtower and the computer interface. Out-of-band channels (such as secure mail and/or SMS) should be established to strengthen the communication redundancy between the watchtower and the wallet owner.

\subsection{Receiving External Payment}

In this process, a wallet owner receives bitcoin from arbitrary transactions (no constraints). ``External" refers to the payment not being subject to constraints by the vault custody protocol (e.g. no requirement for Segregated Witness transaction). For example, this could be a customer transferring bitcoin to a commercial custodian or a custodian sending funds from a separate wallet they own to their active wallet. 

\begin{enumerate}
    \item Custodian validates a receiving address (controlled by the active wallet HMs) using a human-check over an OOB channel for mutual authentication (to thwart address generation attacks).
    \item The computer interface requests an external payment  be sent to the receiving address (communicating with the payee using HTTPS or other secure methods to prevent man-in-the-middle attacks).
    \item External payment is created, signed, and broadcast. It is a transaction with an output that specifies the receiving address. If the payment is large, a penny-test may be used first, or a sequence of fractional payments may be made. 
\end{enumerate}

\subsection{Vaulting}
\label{sec:Vaulting}
This process transfers control of funds from the multi-signature active wallet to a set of \textit{active} vault transactions (where signing keys are deleted and the deposit transaction is confirmed) that are redundantly stored on each of the vault wallet HMs. Moreover, each vault transaction is generated with a corresponding P2RW transaction that will become active. The process makes use of the partially-signed bitcoin transaction (PSBT) format \cite{bip174}. Distributing funds across a set of vault transactions means that accessing a portion (un-vaulting) doesn't reduce the security of all the funds at once.

The following describes how to pass a partially-signed transaction to each of the vault wallet HMs to obtain their signature, to redundantly store the covenant transactions, to ensure that keys are only deleted when the the covenant transaction storage is sufficiently redundant, and to ensure that the computer interface does not ever see in clear-text a partially- or fully-signed vault transaction. This is designed with minimal communication complexity while also minimizing the number of times required to access each device. A covenant transaction requires $m$-of-$n$ signatures and $n-m+1$ ephemeral keys to be deleted. The choice of $m$ determines the fault tolerance to both the compromise and loss of the vault wallet HMs as will be discussed in section \ref{sec:Threat Model}. 

\begin{enumerate}
    \item Vault wallet HMs perform a distributed key generation, each yielding a fresh and ephemeral (private-key, public-key) pair.
    \item The vault wallet HMs communicate the public-keys to each other and construct a consistent $m$-of-$n$ multi-signature target address for the \textit{deposit} and \textit{vault} (recovery path) outputs which is then passed to the computer interface.
    \item The computer interface creates the (Segregated Witness) vault deposit transaction (creating an output which points to the target address) but does not broadcast it. 
    \item The computer interface constructs the vault and P2RW transactions, specifying the; version, locktime, inputs and outputs as defined in section \ref{vault-design}. This should have the format of a partially-signed bitcoin transaction \cite{bip174} with no signatures\footnote{In keeping with the terminology of BIP-174, the computer interface plays the role of the Creator and each of the vault wallet HMs will play the roles of Signer and Combiner}. 
    \item The computer interface requests signatures on the covenant transactions, from $m-1$ of the vault wallet HMs.
    \item The computer interface distributes a copy of the covenant transactions to the remaining $k = n-m+1$ vault wallet HMs, triggering the internal process of each of them to provide the final signature, validate and store the covenant transactions, to delete their signing keys, and to return a notification stating that the ephemeral keys were deleted. Note that the signature is not provided unless the human-check verifies the details of the transactions (assuming secure channels between HMs, this check only needs to occur once). This yields a storage redundancy of $k$.
    \item To increase the storage redundancy, fully-signed covenant transactions are passed to any HMs which don't yet hold them. Each HM validates the transactions, stores them, deletes their ephemeral key and notifies the computer interface of the deletion. At least $n-m+1$ of the HMs must delete their ephemeral key to activate the covenant.
    \item The computer interface requests a {\tt txid}\footnote{The watchtower could also use other unique identifiers for funds such as the hash of the relevant output.} for the vault transaction from one of the vault wallet HMs and sends an authenticated message to the Watchtower to begin watching for a transaction with that {\tt txid}. (A responder watchtower will also receive a copy of the P2RW transaction).
    
    \item The computer interface requests a signature over the deposit transaction from an appropriate threshold of the HMs of the active wallet (with a human-check that the final signature is being applied to the intended transaction). The computer interface then broadcasts the fully-signed deposit transaction. Once the deposit is confirmed, the vault and P2RW covenants are activated. 
\end{enumerate}{}

\subsection{Un-Vaulting}
\label{sec:un-vaulting}
This process describes how the active wallet can re-gain control of the funds which are currently controlled by an active AVT. It is critical that the portion of funds which are un-vaulted at any one time is limited since the security of funds which are un-vaulted reduces to that of the multi-signature active wallet.

\begin{enumerate}
    \item The computer interface requests and obtains a specific AVT with {\tt txid} from one of the vault wallet HMs. 
    \item \textit{(Responder watchtower only)} The computer interface notifies the watchtower nodes with an authenticated message that the AVT with {\tt txid} is going to be broadcast (to protect against accidentally triggering the release of a P2RW transaction).
    \item The computer interface broadcasts the AVT to bitcoin's peer-to-peer network. Once this transaction is included in a block, and once the time-lock expires after that, control of funds is transferred to the active wallet.
    \item The watchtower notifies the wallet owner that an attempt to unvault funds was made. If the amount is too large, a warning is produced.
\end{enumerate}

\subsection{Recovery Processes}

\subsubsection{Recovery from unauthorised un-vaulting}
\label{vault-theft-recovery}
Detecting a theft of AVTs will likely occur when the attacker attempts to broadcast the AVT, but a highly motivated attacker could silently pass the AVT to a miner for direct inclusion in a block without spreading it through the peer-to-peer network. In either case, the associated P2RW transactions should be immediately broadcast, pushing the funds to the recovery wallet HMs. 

With a notification watchtower, at least one of its nodes should notify the wallet owner of the attempted un-vault. The wallet owner will then broadcast the associated P2RW transactions. With a responder watchtower, at least one of the nodes should broadcast the P2RW transaction. In both cases, the fee per \textit{kb} of transaction data  should be significantly high such that the P2RW transaction will confirm before the time-lock expiry of the vault output. 

If the AVTs were all stored on a device that was compromised then the wallet owner should begin to make a \textit{full} recovery by broadcasting each AVT with their associated P2RW transaction. The recovery process also includes transitioning to active use of the recovery wallet and decommissioning the current active wallet. The set-up process should be re-enacted and a new recovery wallet instantiated before using the new active wallet to deposit funds into a new set of vault covenants.  

\subsubsection{Recovery from Active Wallet compromise}

Realistically, an attacker would likely only attempt to un-vault funds if the active wallet has also been compromised (assuming the recovery wallet cannot be compromised) since that is where control of the funds is transferred to. Even then, however, it would make more sense for the attacker to wait for the wallet owner to authorise an un-vault and for the attacker to steal those funds without alerting the wallet owner at all. The theft is limited by the rate of authorised un-vaulting. A theft transaction would likely be detected when the wallet owner attempts to spend the same funds after waiting for the time-lock to expire, and is unable to. Here, the AVTs should be broadcast with their associated P2RW transactions to transfer control to the recovery wallet. The set-up process should be re-enacted, instantiating a new recovery wallet \textit{before} the old recovery wallet becomes the new active wallet.

\subsubsection{Recovery from device failure}

It is likely that a device would fail eventually due to; bitrot, disaster, physical theft, a supply-chain attack, \textit{etc}. The custody protocol is designed to withstand a certain number of device failures through use of threshold signature schemes (see \ref{subsec:Functions and Tolerances}) without custody being effected. 

If an active wallet HM failure is detected, a new replacement device can be rotated into use. In addition to the standard process for setting up a new device (communication channels and key structure) this requires creating a new set of multi-signature addresses (which includes public keys from the new device) to receive external payments. Also new payments (outgoing \textit{or} vault deposits) should prioritise the use of old unspent-transaction-outputs for which the new device cannot provide a signature. 

If a recovery wallet HM failure is detected, a replacement device should be set-up immediately. A new multi-signature address for the recovery wallet should be constructed. Provided the public keys for the recovery wallet HMs were stored on the active wallet HMs as described in \ref{sec:SetUp} then the other recovery wallet HMs don't need to be accessed. All funds controlled by AVTs should be transitioned to a new set of (AVT, P2RW) pairs which specify the new recovery wallet address. To avoid accessing all of the recovery wallet HMs the funds can be un-vaulted and deposited into new vaults in a rate-limited way which doesn't put too much at risk at any one time. 

A vault wallet HM failure would likely be detected during the vaulting process. If this happens, a replacement HM should be set-up and a new vault-wallet multi-signature address should be constructed.  

\subsubsection{Recovery from Privacy loss}

As with all security systems that rely on slow-to-change secrets, the privacy properties of this custody protocol are brittle \cite{BeyondFear}. Once addresses or public keys are leaked for any given wallet, privacy cannot be regained until the control of funds are rotated to a new and \textit{uncorrelated} set of addresses. This would require privacy services that use techniques such as CoinJoin and Mixing \cite{PrivacySurvey2018}, and would require enacting parts of the set-up again to transition to a new key tree for the wallet that was compromised. Of course, any vaulted funds must be un-vaulted (either with the recovery or active wallet) and re-vaulted to accurately specify the new target addresses that are committed to by the covenant transactions. If it is critical for the wallet owner to maintain privacy, then this recovery process severely limits the practicality of this custody protocol. This is because privacy services require on-chain transactions and in the ideal case (providing maximum anonymity) the separate unspent transaction outputs (from the vault transactions) must be mixed through different instances of Mixing or CoinJoin protocols. This means paying for substantial transaction fees and waiting for confirmation times. Methods for improving the privacy properties of the protocol are left for future research, and an interesting place to begin might be the use of off-chain threshold signature schemes rather than on-chain multi-signatures \cite{Gennaro2018,Lindell2018,Canetti2020,Damgard2020,Gennaro2020,Gagol2020}. 

\subsection{Health Check}
\label{subsec:Health Check}

This process checks that all wallets are operational and detects any component failures that must be fixed to re-secure the custody of funds. This process does not require recovery wallet HMs (or any set of HMs) to be brought to a single location. The goals for a health check are similar to the goals of a proof-of-assets (often used by exchanges in conjunction with a proof-of-liabilities as part of a proof-of-solvency \cite{Provisions, Decker2015MakingBE}). Proving that one has control of some funds corresponds to demonstrating one can access a UTxO with a script and set of signatures that will be validated according to the consensus protocol. 

The protocol for demonstrating that devices in the active and recovery wallets are healthy works by demonstrating control of the master extended private key (the key at the root of the key-tree and the chaincode used to derive child keys \cite{BIP32}). This shows that each device has access to (either directly or through derivation) it's entire key-tree. This can be done as per BIP 127 \cite{BIP127}, a standard for proof-of-reserves that is easy to integrate with existing wallet infrastructure since it relies on signing an invalid transaction and uses the PSBT format \cite{bip174}. Provers must be careful to avoid biased nonce generation \cite{Breitner2019} and nonce reuse to retain security of their ECDSA signatures. This isn't privacy preserving, but it doesn't need to be if the message signed is a signature hash which commits to an invalid transaction, since knowledge of the signature can't be used to steal funds. 

A different protocol is needed for demonstrating that the vault wallet devices have securely stored the AVTs. Since the private keys have been deleted, they cannot be used to create new signatures, and can not be used to demonstrate access to funds. Instead, the vault HMs need to provide a zero-knowledge proof-of-possession of a set of signatures \cite{ZKPoPDS} and script for a valid transaction which spends a specific UTxO without revealing those signatures. If the signatures are revealed to the device performing the health check, the security of storage is significantly reduced. Revealing the signature to a compromised device would enable an early broadcast of the vault transaction. Efficient proofs-of-knowledge for algebraic statements (such as the verification of an ECDSA signature) can be generated using a class of proof systems known as Sigma protocols \cite{NIZKP2018}, where the statement (or \textit{relation}) to be proved is decomposed into a circuit of gates \cite{ZKP2016}. Sigma protocols can be made non-interactive with the Fiat-Shamir transform \cite{FiatShamir}. To the best of the authors' knowledge there is not yet an implementation that would enable such proofs. This would be an important contribution for bitcoin protocols that make use of pre-signed transactions with secure key deletion.

In addition, the watchtowers and each communication channel with them are checked. Communications can be checked with a simple heart-beat request. The watchtower should demonstrate that it has correct knowledge of the set of UTxOs to watch. In the case were a \textit{responder} watchtower is in use, it must demonstrate that it has access to the relevant set of P2RW transactions. This amounts to a simple consistency check. No privacy preservation is required between the computer interface and the watchtower since both already have access to this information. 

\section{Threat Model}
\label{sec:Threat Model}

This threat model assumes a correct set-up and that human operators act honestly and correctly during the enactment of each process.  A security analysis for malicious internal threats remains an open question for future research. This threat model assumes the wallet owner does not require maintenance of privacy for the addresses they use.

To understand the myriad ways in which the vault custody protocol can go wrong and to what extent funds can be lost or stolen in each scenario, one must first have a clear understanding of the tolerances of individual functions of the custody protocol. Then, one must consider the optimal strategy of an attacker who successfully compromises a functionality or multiple functionalities together. A configuration specifies the number of devices/ machines and which subsets of them perform each function. This discussion  is independent of the specific configuration of devices, and will inform how different options  affect the security properties offered by the custody protocol as a whole. It is critical to determine where independence is required between functionalities to compartmentalize risk. 

\subsection{Functions and Tolerances}
\label{subsec:Functions and Tolerances}

\vspace{0.1cm}
\textbf{\noindent Recovery Wallet:} The security of the recovery wallet functionality can be characterized as a standard $m-$of$-n$ multi-signature wallet. Of the $n$ signing keys that are stored on separate, geographically distributed, off-line HMs, the functionality resists up to $n-m$ arbitrary failures in which signing keys are lost and resists arbitrary attacks where up to $m-1$ of the signing keys are leaked.

\vspace{0.1cm}
\noindent \textbf{Active Wallet:} The security of the active wallet functionality can be characterized as a standard $j-$of$-k$ multi-signature wallet. Of the $k$ signing keys that are stored on separate, off-line HMs, the functionality resists up to $k-j$ arbitrary failures in which signing keys are lost and resists arbitrary attacks where up to $j-1$ of the signing keys are leaked.

\vspace{0.1cm}
\noindent \textbf{Secure Key Deletion:} The preparation and activation of the two types of covenant transactions (vault transactions and push-to-recovery-wallet transactions) occurs on $p-$of$-t$ off-line HMs. This functionality can resist up to $p$ compromises where the deleted signing key for the covenant is recovered (through malware, side-channel attacks, or physical inspection attacks). 

\vspace{0.1cm}
\noindent \textbf{Vault Transaction Storage:} There are two possible failure modes for this functionality; loss and theft of transactions. Assume that the vault transactions are stored on $R$ storage devices. This functionality resists loss with up to $R-1$ arbitrary device failures. This functionality is compromised by theft with a single arbitrary malicious compromise of a storage device. 

\vspace{0.1cm}
\noindent \textbf{P2RW Transaction Storage:} There are two possible failure modes for this functionality; loss and theft of transactions. Assume that the P2RW transactions are stored on $S$ storage devices. This functionality resists loss with up to $S-1$ arbitrary device failures. This functionality is compromised by theft with a single arbitrary malicious compromise of a storage device. 

\vspace{0.1cm}
\noindent \textbf{Notification Watchtower:} The watchtower functions as a simple notification provider that alerts the wallet owner of any attempts to un-vault funds it is guarding. For this, the watchtower must maintain up-time, bitcoin network connectivity, and secure communication channels with the wallet owner. This functionality is best achieved with a distributed set of full bitcoin nodes. Assuming there are $W$ watchtower nodes, then this functionality resists at most $W-1$ arbitrary node failures.

\vspace{0.1cm}
\noindent \textbf{Responder Watchtower:} The responder watchtower operates as a notification provider (as above), and is also responsible for triggering the recovery process (by broadcasting P2RW transactions) when it detects an un-authorized un-vault. In addition this watchtower must securely (robustly and privately) store the push-to-recovery-wallet transactions and resist being bribed or fooled into releasing them un-necessarily. This functionality is a coupling of P2RW Transaction Storage and the Notification Watchtower, and is brittle since the leakage of the set of P2RW transactions from a single node enables denial-of-service to a wallet owner attempting to un-vault funds. Observe however that this doesn't enable theft of funds.

\vspace{0.1cm}
\noindent \textbf{Fee Wallet:} The fee wallet functionality is performed either by a responder watchtower, or the wallet owner (when a notification watchtower is used). In the former case, this functionality is thus only as secure as the watchtower which must also maintain wallet security (privacy and key confidentiality). In the latter case, the fee wallet functionality can be characterised as a standard $a-$of$-b$ multi-signature wallet. Of the $b$ signing keys that are stored on separate, off-line hardware modules, the functionality resists up to $b-a$ arbitrary failures in which signing keys are lost and resists arbitrary attacks where up to $a-1$ of the signing keys are leaked.

\vspace{0.1cm}
\noindent \textbf{Human-Check:} This functionality protects against payment and address generation attacks, and remains secure as long as at least one communication channel (in-band or out-of-band) between the HMs and the human remain secure.

\vspace{0.1cm}
\noindent \textbf{Computer Interface:} It is assumed that the computer interface is corruptible. 

\subsection{Adversaries}

There are numerous expected adversaries that this custody protocol can tolerate to a certain degree through resilient design (redundant and diversely implemented functionalities).

The HMs may be corrupted upon purchase (e.g. malicious firmware, evil-maid attack), may be physically compromised, may be subjected to buffer-overflow and speculative execution attacks, may suffer bit-rot, or may be lost or destroyed in a disaster. The computer interface may be corrupted, enabling the following attacks when interacting with a HM \cite{FormalHardware}:
\begin{enumerate}
    \item[] \textit{Payment attack}. During transaction issuing, an attacker may tamper with inputs, receiving addresses and fees. This is guarded against with human check (see \ref{sec:OOBMutAuth}).
    \item[] \textit{Address generation attack}. An attacker generates receiving addresses for client that is controlled by the attacker. This is guarded against with human check (see \ref{sec:OOBMutAuth}).
    \item[] \textit{Privacy loss}. When an attacker corrupts the computer interface, they can access public keys, addresses and balances that are loaded into memory, but these are not persistently stored.
    \item[] \textit{Chain attack}. An attacker tampers with balance calculations by providing a malicious chain to the computer interface. Given that the watchtower is designed to be resilient to such attacks, this attack can be mitigated through consistency checks with the watchtower nodes. 
\end{enumerate}

The human operator may face denial of access to some of their HMs if the physical security of the device is dependent on another service/ person. The human operator may be coerced by violence to access funds for a targeted theft or legal forfeiture. In these circumstances, a hidden trigger should be implemented for the recovery process such that funds are pushed to the geographically distributed recovery wallet. 
The following are adversaries that are not handled by this custody protocol. This protocol does not protect against death or sudden disability. For that, a `dead-man switch' mechanism could be introduced to pass on control of funds to an heir, but this is work for future research. Any loss of fungibility through correlation and blacklisting of funds should be handled before transitioning the custody of funds to this protocol. Failures by the human operator due to social engineering, process fatigue, and general errors are not protected with this custody protocol. 

\subsection{Catastrophic Failure Modes}
\label{subsec:Catastrophic}

To illustrate how the benefits drawn from the vault custody protocol depend on a secure recovery process, the impact of a compromised recovery wallet will now be discussed in conjunction with other compromised functionalities.

\begin{enumerate}
    \item \textit{Recovery wallet compromise:} An attacker who successfully compromises a threshold of signing keys from the recovery wallet (despite the stringent procedural limitations for their accessibility) can (while the wallet owner is unaware) easily claim any funds that are ‘recovered’, after the broadcast of P2RW transactions. The optimal strategy for an attacker here would be to wait for an attempt to recover funds (forced through some other attack or loss scenario) and to claim those funds. In response, once notified of the breach, the wallet owner understands that they should keep the remaining P2RW transactions private and only access funds through the time-locked spending path of the AVTs. The loss incurred here will be limited by the rate that P2RW transactions are broadcast.
    
    \item \textit{Recovery wallet and P2RW compromise:} An attacker who successfully compromises a sufficient threshold of private keys from the recovery wallet and gains access to the full set of P2RW transactions will have knowledge of the distribution of funds across AVTs. Since they cannot force the wallet owner to attempt to un-vault funds, their best strategy would be to wait for the AVT with a large portion of funds to be broadcast, and to claim it by broadcasting the associated P2RW transaction with a theft transaction.
    If the breach is detected, the wallet owner could choose to thwart further thefts by keeping the remaining AVTs private and freezing those funds. Alternatively, the wallet owner could attempt to race the attacker by creating transactions with higher priority (or by collaborating directly with a miner). The theft here is limited by the rate of broadcast of AVTs but the loss could be total (if the wallet owner decides to freeze all funds).

    \item \textit{Recovery wallet and AVT compromise:} In the case where an attacker successfully compromises a sufficient threshold of keys from the recovery wallet and the set of AVTs, the strategy space for attack is complicated and depends upon the response of the wallet owner. Assuming that the wallet owner is unaware of the compromise of the recovery wallet but is notified by the watchtower that the set of AVTs has been broadcast by an attacker, the wallet owner would attempt to recover funds by broadcasting the P2RW transactions (since allowing the time-locks to expire would reduce the security of all funds to that of the active wallet, which is supposed to be less well secured than the recovery wallet). This would enable the attacker to broadcast theft transactions and attempt to claim all funds. 
    If instead the wallet owner detects or suspects that the recovery wallet is compromised too, they may thwart the attack by keeping the P2RW transactions private and allowing the time-locks to expire, perhaps re-vaulting funds with a new set of recovery devices. This demonstrates a motivation for limiting the rate of release of P2RW transactions when a compromise is suspected. 
    
    \item \textit{Recovery wallet, AVT and P2RW compromise:} An attacker in this case can trigger total recovery of funds by broadcasting all AVTs and P2RW transactions, claiming those funds with theft transactions which evade the time-locked spending paths. A vigilant and swift wallet owner may be able to salvage some of the funds if they can generate transactions that recover funds with higher priority than the attackers' theft transactions. The amount lost or stolen depends on the ability of the wallet owner to have their transactions mined before the attacker’s theft transactions and ranges from small loss (limited by the amount stolen before the intrusion detection) to total loss of funds. This demonstrates why the storage of covenant transactions should be totally independent from the recovery wallet. Independent storage of AVTs from P2RW transactions also improves resistance to this failure mode.
    
    \item \textit{Recovery wallet and watchtower compromise:} One might consider ways in which the attacker could convince the wallet owner that the AVTs or active wallet were compromised, in order to trigger a recovery process (broadcasting the P2RW transactions), enabling the attacker to claim all funds with theft transactions. This could be done by maliciously triggering the watchtower with a fake un-vault transaction (a watchtower should be designed to validate un-vault transactions to avoid this). Perhaps there are other ways to attack the watchtower, such as an eclipse attack that convinces the watchtower that an un-vault transaction has occurred (a watchtower should consist of at least two separate nodes with diverse connections). Perhaps a m-i-t-m attack can compromise the communications channel between the watchtower and the wallet owner (a watchtower should have multiple out-of-band communication channels with the wallet owner). If the watchtower is not secure, then this is the worst case scenario and a total loss of funds could occur. Provided at least one of the watchtower machines remains secure and the intrusion is detected, the wallet owner would have to race the attacker by broadcasting recovery transactions with higher priority than the theft transactions in an attempt to salvage some of the losses.
    
\end{enumerate}{}

\noindent Other catastrophic failure modes which don't involve a compromised recovery wallet include; 

\begin{enumerate}
    \item[6.] \textit{Active wallet and AVT and watchtower compromise:} Consider an attacker who successfully compromises the active wallet signing keys and gains access to the AVTs and manages to stop the watchtower from notifying the wallet owner of attempts to un-vault funds. The attacker can freely broadcast the AVTs, wait for the time-locks to expire, and spend funds using the active wallet's signing keys. This catastrophic failure could incur a loss of all funds. 
    
    \item[7.] \textit{Failed key deletion/ ephemeral vault key compromise:} An attacker who successfully compromises a sufficient threshold of signing keys and addresses from the vault wallet (either during their lifecycle or by recovery after deletion) could revoke the covenants which control funds and by-pass their constraints. A set of theft transactions could be created which spend all funds that were thought to be controlled by AVTs. The wallet owner may be alerted of this provided the watchtower can detect attempts to spend from vault deposit outputs. Then the wallet owner would be in a race to have the AVTs mined before the theft transactions.
    
    \item[8.] \textit{Failed Key deletion/ ephemeral vault key and watchtower compromise:} An attacker who successfully compromises a sufficient threshold of signing keys and addresses from the vault wallet (either during their lifecycle or by recovery after deletion) and gains control of the watchtower can broadcast arbitrary theft transactions without the wallet owner being notified. This scenario is the worst case, resulting in a total loss of vaulted funds. 
    %
    
\end{enumerate}{}

\subsection{Limited Loss Scenarios}
\label{subsec:LimitsOfLoss}

Assume for the following argument that the recovery process (transferring control of funds to the recovery wallet) is secure; it can be performed within the delay period given by the relative time-locks of the AVTs, and a sufficient threshold of recovery wallet HMs remain un-compromised.

Before discussing various strategies that could be used by an attacker, it is worth discussing two aspects of user policy which will limit losses in the event of a significant compromise. The first aspect is the distribution of funds across multiple AVTs. The second is the acceptable rate (volume per time interval) of funds to be un-vaulted. There should be well-defined policies for each aspect in order to appropriately manage the risk of funds in custody. 

Consider a scenario where all funds are controlled by a single AVT. When the wallet owner un-vaults these funds, once the time-lock expires, the security of those funds reduces to that of the active wallet. This demonstrates the need for partitioning funds across multiple AVTs, and un-vaulting funds at a limited rate in order to retain the security afforded by the recovery process for the majority of funds while the security of only a small portion is reduced to that of the active wallet.

Thus, there are two parameters for tuning the risk-management. The first is the distribution profile of vaulted funds and the second is the acceptable rate of un-vaulting. While distributing funds thinly across many AVTs enables a slow rate of un-vaulting, it increases the transaction fee burden, and the procedural burden of creating AVTs. The rate of un-vaulting determines the maximum loss of funds for various scenarios of compromise as discussed below. The optimal strategy of an attacker who successfully compromises different functions of the vault custody protocol (secure AVT storage, active wallet, watchtower) in various combinations will be shown along with the loss incurred in each situation.

\begin{enumerate}
    \item \textit{Theft of AVTs:} An attacker gains control of a device which controls a set of AVTs but is unable to compromise a sufficient threshold of signing keys for either the active wallet or the recovery wallet. First, there is a loss of privacy associated with loss of AVTs that may make the whole custody protocol easier to attack. The worst an attacker can do is to broadcast these transactions, and (assuming secure watchtower) force the wallet owner to go through the recovery process. Alternatively, the attacker may wait until they successfully compromise the active wallet too, in which case the knowledge of the details of the AVTs will yield a more profitable attack.
    
    \item \textit{Active wallet compromise:} An attacker who successfully breaches the active wallet can attempt to gain unrestricted control over funds by waiting for them to be un-vaulted (so as not to alert the wallet owner). An unwitting wallet owner will un-vault funds and wait for the time-lock to expire. The attacker must then have the theft transaction be mined before the wallet owner spends those funds as intended.  If the wallet owner adheres to a low un-vault rate policy, the maximum loss in this scenario is limited, and the breach will be noticed when the wallet owner fails to spend those funds, enabling a full recovery of the rest of funds.  
    
    \item \textit{Active wallet and watchtower compromise:} Given that the watchtower is programmed to detect un-vaulting attempts, but isn't aware of the exact details of the AVTs nor of the intended, subsequent, non-fraudulent transactions from the active wallet, it is not beneficial for an attacker to compromise the watchtower (at least without also stealing some AVTs). An attacker who successfully compromises a threshold of signing keys from the active wallet gains nothing from also compromising the watchtower because the wallet owner is alerted to theft transactions when attempting to spend the un-vaulted funds, not by the watchtower. 
    
    \item \textit{Active wallet and AVT compromise:} An attacker who successfully compromises a threshold of signing keys from the active wallet and breaches the devices which are storing the AVTs has a choice of strategies. The attacker could broadcast any or all AVTs and attempt to claim them by waiting for the time-locks to expire, but this would alert the wallet owner and so is not likely. The attacker can instead wait, and knowing the distribution of funds across AVTs, can decide to attack when it is most profitable. As soon as the theft transaction which is passed to the miners (optimally not through broadcast, but privately) becomes public knowledge (enters the mempool or is mined), the attacker can assume the wallet owner has been alerted and broadcast all remaining AVTs, attempting to overwhelm the wallet owner's recovery process. The wallet owner (assuming secure watchtower and effective time-lock lengths) will recover these funds, limiting loss to the largest partition of funds.
    
    \item \textit{Watchtower and AVT compromise:} An attacker who successfully compromises the watchtower and gains access to a device which stores AVTs can cause grief for the wallet owner by broadcasting the AVTs and shutting down the watchtower so that the wallet owner is not notified. Once the time-locks expire, the security of those funds reduces to that of the active wallet (or recovery wallet) but there is no loss of funds. There will also be a loss of privacy associated with the loss of AVTs that can be used against the wallet owner.
    
    \item \textit{Theft of P2RW transactions: } An attacker who gains access to the set of P2RW transactions will gain some information about the distribution of vaulted funds. They will also be able to force the wallet owner to access their recovery wallet if an attempt to un-vault funds is made, though the wallet owner will be alerted when attempting to spend those funds (or by the watchtower). The rate of broadcast of AVTs will determine the amount of funds that the attacker can effect this way. So long as the wallet owner maintains redundant storage of P2RW transactions, no loss of funds will be incurred without additional functionality compromise.  
    
    \item \textit{Watchtower compromise: } An attacker who compromises the watchtower will gain access to some sensitive information (addresses, covenant transaction IDs, communication channel information) and will be able to tamper with notifications which will make attacking other functionalities easier. No loss of funds will be incurred without additional functionality compromise.
    
    \item \textit{Watchtower and P2RW transaction compromise: } Given control of the watchtower and theft of the set of P2RW transactions, an attacker can tamper with operations and break privacy in the ways described above. No direct loss of funds will be incurred without additional functionality compromise. 
    
    \item \textit{Human-check OOB channel compromise: } An attacker who successfully compromises the in-band and OOB communication channel used to add redundancy to the human-check for each use of a HM will be able to perform address generation and payment attacks. This could mean disrupting vault deposits, vault transactions, P2RW transactions, and arbitrary transactions from the active wallet.
    
    \item \textit{Fee wallet compromise: } If fees are to be dynamically allocated to ensure the right priority when broadcasting covenant transactions, an attacker who compromises the fee wallet can perhaps hinder the ability to update covenant transactions appropriately. Thus, an attacker who successfully compromises a sufficient threshold of signing keys from the fee wallet can not only steal the funds controlled by the fee wallet, but can also disrupt attempts to un-vault funds (or push funds to the recovery wallet). A compromised fee wallet could be a serious hindrance in any scenario where the wallet owner must race against an attacker's theft transactions. 
    %
    
\end{enumerate}{}

\subsection{No Loss Scenarios}

Any scenario where no individual function is compromised (i.e. where the tolerance threshold is not breached) will not incur direct loss of funds, however, privacy may be lost and operational costs incurred (e.g. from new hardware or requiring re-enacting parts of the set-up process). 

\section{Script-based Covenants}
\label{Sec:ScriptCovenants}

A recent BIP (the latest in a series of proposals) presents a Script-based mechanism for covenants in bitcoin by introducing a new OP\_CODE called OP\_CHECKTEMPLATEVERIFY (CTV) \cite{BIP119}. An implementation is available on github \cite{CTVImplementation}. Whether or not this alternative covenant mechanism would enable less cumbersome designs for vault-custody protocols is an important argument for the progress of this soft-fork upgrade. A vault-custody protocol will thus be outlined which uses CTV instead of the pre-signed transaction with secure key deletion mechanism.

CTV scripts commit funds in an unspent transaction output to be spent by a transaction with a specified set of values for the serialized version, locktime, scriptSigs hash (if any non-null scriptSigs), number of inputs, sequences hash, number of outputs, outputs hash, and currently executing input index. A commitment hash is produced with these values and is included in the output of the transaction that deposits bitcoin into the covenant. To enforce the vault logic, a specific address which is constructed from a vault locking script (as in \ref{vault-design}) would be committed to. 

The proposed architecture shown in figure \ref{fig:Architecture} would no longer need dedicated devices for the construction and endorsement of vault and push-to-recovery-wallet covenants. The covenants, instead, would be enforced by the bitcoin protocol. The addresses for vault and P2RW transactions can be generated well in advance since there is no need to minimise the life-cycle of ephemeral keys. Because of this, covenant activation is non-interactive, a property which lends itself better to multi-party protocols. The vaulting process would thus be simplified along with reduced procedural overhead for device set-up and management. To make-use of the enforced time-lock, CTV vault-custody would still require a watchtower. Similarly, for defence-in-depth and minimized network exposure, multi-signature active and recovery wallets with numerous off-line HMs would still be required.

The transaction templates that encode the desired constraints of a CTV covenant must be planned in advanced and stored securely. The planning is non-interactive and deterministic. The architecture for deleted-key covenant vault-custody has an obvious choice of storage device, the vault wallet HMs, since they are generally disconnected from networked devices and already handle the creation of the covenant transactions. However, with CTV-based vault-custody, there is no vault wallet and storage must occur elsewhere. Storage of planned transaction templates should be confidential and redundant. As with deleted-key vault-custody, theft of vault transactions can force the wallet owner into the recovery process in a DoS type of attack. Since CTV has significantly improved set-up, if this attack is critical to defend against (e.g. if the custody protocol is used by active traders) then it may be practically feasible to distribute vault transactions with a verifiable secret sharing scheme \cite{VSS}. Storing the planned transactions independently from the active wallet HMs is optimal since then independent barriers are posed to an attacker rather than compounding the risk by requiring the same set of devices to secure the active wallet and planned transaction storage functionalities (recall scenario 6 in section \ref{subsec:Catastrophic}). It is critical that all planned transactions include some entropy to ensure they aren't easily derivable with minimal information about the public-keys in use and with reasonable guesses about the access policy.

For CTV vault-custody, the method for dynamically increasing transaction fees is improved when compared with deleted-key covenants. The basic mechanism is to write output scripts that conditionally commit to two variants of the planned transactions. For example, commit to planned transaction $x$ \textit{or} $x'$, where $x'$ has an additional input that is used to increase the fee. The output script would have the form 

\begin{verbatim}
          OP_IF <Hash(x)> OP_ELSE <Hash(x')> OP_ENDIF OP_CTV  
\end{verbatim}{}

\noindent where {\tt Hash(transaction)} represents the hash from the specification in BIP-119 \cite{BIP119}. Critically, if Segregated Witness addresses are used, the additional input in transaction $x'$ doesn't need to be known or prepared in advanced. For Segregated Witness transactions the signatures are not committed to by {\tt Hash(transaction)} because the scriptSigs are empty. In fact, a chain of covenant transactions can be planned which each commit to two variants of the subsequent transaction, where one variant has an additional input for paying a fee. This would mean that funds bound by vault covenants can be spent by a vault transaction whose fee can be dynamically adjusted without rendering the P2RW transaction invalid. While CPFP wouldn't be needed, it will still be available. Moreover, transactions that spend from CTV-bound outputs can be replaced using RBF if needed.

One important task for custody operations in some business contexts is to prove to an auditor that one has access to the funds, and that those funds are bound by a specific type of covenant. With deleted-key covenants, proving that funds are accessible would require a zero-knowledge proof-of-possession of a digital signature for a valid transaction (as discussed in section \ref{subsec:Health Check}). Proving that a covenant has been enforced requires the participation of the auditor during the set-up. This can be achieved by having the auditor participate in the key-deletion process by giving them sufficient signing keys ($t-p+1$) to enforce the covenant themselves. With CTV-based covenants it is far simpler to prove that one has access to funds and that they are bound by a covenant. The former proof can be done by signing an invalid transaction with the appropriate private key, as per the proof-of-reserves protocol \cite{BIP127}. The latter can be proved by inspecting the covenant-bound unspent transaction output that has been confirmed on the public blockchain, and sharing the planned transaction tree that was committed to with the auditor.  

While this overview is not comprehensive of the capabilities and risks of a CTV soft-fork upgrade, it demonstrates that custody protocols that rely on covenants may be simplified significantly when compared with what is currently possible in bitcoin. An ability to enforce automated transaction flows while exposing a minimal attack surface can benefit self-custody and joint-custody protocols by acting as choke points. Reducing the procedural overhead and interactivity requirements by using CTV  offers a clear advantage to the design, implementation and operation of covenant-based custody protocols.

\section{Conclusion}

A custody protocol has been proposed as a means for an owner of bitcoin to protect their funds from loss and theft. There appears to be a trade-off between protection from loss and protection from theft. The former requires redundancy and distributed storage which increases the attack surface for the latter. Moreover, while a diverse multi-component custody operation increases the cost of attack, it also increases the procedural complexity for the wallet owner and thus increases the risk of accidentally losing access to funds. The use case discussed assumed no malicious insider threats; any operators of the protocol are assumed to act honestly. However, the protocol may be useful in multi-party contexts if there are legal or physical security means for maintaining accountability and thus minimizing malicious insider behaviour. The custody protocol protects against arbitrary external threats up to a (specifiable) threshold number of failures/ compromises for different functions of the protocol. Under the most likely theft scenarios, the loss of funds is limited to a pre-defined portion of funds through compartmentalization and rate-limiting. 

The proposed solution has been designed to demonstrate the utility of vault and push-to-recovery-wallet covenant transactions as a mechanism for restricting access to funds with time-locks and enabling a recovery process if an intrusion is detected or suspected. Moreover, the general requirements for integrating vault covenants into a custody protocol have been demonstrated to be an active wallet, a recovery wallet, and a watchtower. An additional vault wallet is only necessary when implementing covenants using pre-signed transactions with secure key-deletion. The security of funds that are held by vault covenants (the majority) is dependent on: the correct functioning of the proof-of-work consensus protocol of bitcoin for enforcement of time-locks and multi-signature access policies, on the process of secure key deletion to enforce restricted spending patterns with covenants, and on the watchtower for enabling recovery when a compromise is detected.

This protocol affords defence-in-depth to the custody of bitcoin, as with multi-signature wallets. The additional value provided by this protocol is the ability to specify and tune the architectural parameters to achieve a desirable balance for an inherent security-accessibility trade-off. For example, increasing the length of the time-lock reduces convenience (funds can't be accessed as quickly) but increases security of the recovery process. Another adjustable parameter is the rate of un-vaulting; reducing the amount of funds that are accessed per unit time is inconvenient but it affords better security by putting less at risk at any one time. Finally, the number of devices used to yield fault tolerance for each functionality is also adjustable, and generally more devices would result in less convenience but more security. 

The custody protocol has complicated set of processes that require human participation and requires management of numerous hardware devices. The lack of simplicity means that it is not a worthwhile approach for small amounts of funds, or for someone without a certain level of expertise or training. However, for large amounts of funds and sufficient expertise the protocol offers flexibility with risk-management and accessibility that a conventional multi-signature wallet approach would not.

Finally, an outline is given of how this custody protocol (and covenant-based custody protocols in general) can be simplified with use of a script-based covenant mechanism. The simplification is due to bypassing the need for a secure key deletion process which simultaneously tightens the security assumptions and reduces procedural overhead of enacting the custody processes. This is compelling evidence for the benefits of a soft-fork upgrade to bitcoin such as BIP-119 \cite{BIP119}. 

\section*{Acknowledgements}

We thank Professor McBurney (King's College London), Sam Abbassi (Fidelity Center for Applied Technology), Antoine Poinsot (Leonod) and Jeremy Rubin for constructive feedback on this work. We thank Stepan Snigirev for discussion and guidance on creating the prototype custom hardware modules.

\section*{Funding}

Funding is gratefully acknowledged under a UK EPSRC-funded GTA Award through King's College London, and from EPSRC Research Grant EP/P031811/1, the Voting Over Ledger Technologies (VOLT) Project. This work was also supported by Fidelity Center for Applied Technology. 

\bibliography{sample}

\newcommand{\etalchar}[1]{$^{#1}$}
\begin{thebibliography}{TWMP05}

\bibitem[AGKK19]{FormalHardware}
Myrto Arapinis, Andriana Gkaniatsou, Dimitris Karakostas, and Aggelos Kiayias.
\newblock A formal treatment of hardware wallets.
\newblock In {\em Financial Cryptography and Data Security}, pages 426--445.
  Springer International Publishing, 2019.

\bibitem[AGM18]{NIZKP2018}
Shashank Agrawal, Chaya Ganesh, and Payman Mohassel.
\newblock Non-interactive zero-knowledge proofs for composite statements.
\newblock In {\em Advances in Cryptology -- CRYPTO 2018}, pages 643--673.
  Springer International Publishing, 2018.

\bibitem[BH19]{Breitner2019}
Joachim Breitner and Nadia Heninger.
\newblock Biased nonce sense: Lattice attacks against weak {ECDSA} signatures
  in cryptocurrencies.
\newblock Cryptology ePrint Archive, Report 2019/023, 2019.

\bibitem[CGM16]{ZKP2016}
Melissa Chase, Chaya Ganesh, and Payman Mohassel.
\newblock Efficient zero-knowledge proof of algebraic and non-algebraic
  statements with applications to privacy preserving credentials.
\newblock In {\em Advances in Cryptology -- CRYPTO 2016}, pages 499--530.
  Springer Berlin Heidelberg, 2016.

\bibitem[CGMA85]{VSS}
B.~{Chor}, S.~{Goldwasser}, S.~{Micali}, and B.~{Awerbuch}.
\newblock Verifiable secret sharing and achieving simultaneity in the presence
  of faults.
\newblock In {\em 26th Annual Symposium on Foundations of Computer Science
  (sfcs 1985)}, pages 383--395, Oct 1985.

\bibitem[Cho17]{bip174}
Andrew Chow.
\newblock Partially signed bitcoin transaction format, 2017.
\newblock \url{https://github.com/bitcoin/bips/blob/master/bip-0174.mediawiki}
  (18 May 2020, Last Accessed).

\bibitem[CMP20]{Canetti2020}
Ran Canetti, Nikolaos Makriyannis, and Udi Peled.
\newblock Non-interactive, proactive, threshold {ECDSA}.
\newblock Cryptology ePrint Archive, Report 2020/492, 2020.

\bibitem[DBB{\etalchar{+}}15]{Provisions}
Gaby~G. Dagher, Benedikt B\"{u}nz, Joseph Bonneau, Jeremy Clark, and Dan Boneh.
\newblock Provisions: Privacy-preserving proofs of solvency for bitcoin
  exchanges.
\newblock In {\em Proceedings of the 22nd ACM SIGSAC Conference on Computer and
  Communications Security}, CCS ’15, page 720–731. Association for
  Computing Machinery, 2015.

\bibitem[DGSW15]{Decker2015MakingBE}
Christian Decker, James Guthrie, Jochen Seidel, and Roger Wattenhofer.
\newblock Making bitcoin exchanges transparent.
\newblock In {\em European Symposium on Research in Computer Security}, 2015.

\bibitem[DJN{\etalchar{+}}20]{Damgard2020}
Ivan Damgård, Thomas~Pelle Jakobsen, Jesper~Buus Nielsen, Jakob~Illeborg
  Pagter, and Michael~Bæksvang Østergård.
\newblock Fast threshold {ECDSA} with honest majority.
\newblock Cryptology ePrint Archive, Report 2020/501, 2020.

\bibitem[ET05]{PSKTLS}
P.~Eronen and H.~Tschofenig.
\newblock {Pre-Shared Key Ciphersuites for Transport Layer Security (TLS)},
  2005.
\newblock \url{https://tools.ietf.org/html/rfc4279}.

\bibitem[FBDk15]{BIP68}
Mark Friedenbach, BtcDrak, Nicolas Dorier, and kinoshitajona.
\newblock Relative lock-time using consensus-enforced sequence numbers, 2015.
\newblock \url{https://github.com/bitcoin/bips/blob/master/bip-0068.mediawiki}
  (18 May 2020, Last Accessed).

\bibitem[FS87]{FiatShamir}
Amos Fiat and Adi Shamir.
\newblock How to prove yourself: Practical solutions to identification and
  signature problems.
\newblock In {\em Proceedings on Advances in Cryptology---CRYPTO ’86}, page
  186–194. Springer-Verlag, 1987.

\bibitem[GG18]{Gennaro2018}
Rosario Gennaro and Steven Goldfeder.
\newblock Fast multiparty threshold ecdsa with fast trustless setup.
\newblock In {\em Proceedings of the 2018 ACM SIGSAC Conference on Computer and
  Communications Security}, CCS '18, pages 1179--1194, New York, NY, USA, 2018.
  ACM.

\bibitem[GG20]{Gennaro2020}
Rosario Gennaro and Steven Goldfeder.
\newblock One round threshold {ECDSA} with identifiable abort.
\newblock Cryptology ePrint Archive, Report 2020/540, 2020.
\newblock \url{https://eprint.iacr.org/2020/540} (19 May 2020, Last Accessed).

\bibitem[GKSS20]{Gagol2020}
Adam Gagol, Jedrzej Kula, Damian Straszak, and Michal Swietek.
\newblock Threshold {ECDSA} for decentralized asset custody.
\newblock Cryptology ePrint Archive, Report 2020/498, 2020.

\bibitem[Hol16]{AntiFragileICT}
Kjell Hole.
\newblock {\em Anti-fragile ICT Systems}.
\newblock Springer International Publishing, 2016.

\bibitem[KL18]{PrivacySurvey2018}
M.~C. {Kus Khalilov} and A.~{Levi}.
\newblock A survey on anonymity and privacy in bitcoin-like digital cash
  systems.
\newblock {\em IEEE Communications Surveys Tutorials}, 2018.

\bibitem[LN18]{Lindell2018}
Yehuda Lindell and Ariel Nof.
\newblock Fast secure multiparty ecdsa with practical distributed key
  generation and applications to cryptocurrency custody.
\newblock In {\em Proceedings of the 2018 ACM SIGSAC Conference on Computer and
  Communications Security}, {CCS '18}, pages 1837--1854, New York, NY, USA,
  2018. ACM.

\bibitem[LSA20]{Latvala2019}
Sampsa Latvala, Mohit Sethi, and Tuomas Aura.
\newblock Evaluation of out-of-band channels for iot security.
\newblock {\em SN Computer Science}, 2020.

\bibitem[McE16]{P2TST}
Bob McElrath.
\newblock Re-imagining cold storage with timelocks.
\newblock 2016.
\newblock
  \url{https://medium.com/@BobMcElrath/re-imagining-cold-storage-with-timelocks-1f293bfe421f}
  (18 May 2020, Last Accessed).

\bibitem[MES16]{moeser2016bitcoin}
Malte M{\"o}ser, Ittay Eyal, and Emin~G{\"u}n Sirer.
\newblock Bitcoin covenants.
\newblock In {\em FC '16: Proceedings of the the 20th International Conference
  on Financial Cryptography}, 2016.

\bibitem[NBMV99]{ZKPoPDS}
Khanh~Quoc Nguyen, Feng Bao, Yi~Mu, and Vijay Varadharajan.
\newblock Zero-knowledge proofs of possession of digital signatures and its
  applications.
\newblock In {\em Information and Communication Security}, pages 103--118.
  Springer Berlin Heidelberg, 1999.

\bibitem[New18]{Pinning}
John Newbery.
\newblock What is meant by transaction `pinning'?, 2018.
\newblock
  \url{https://bitcoin.stackexchange.com/questions/80803/what-is-meant-by-transaction-pinning/80804#80804}
  (18 May 2020, Last Accessed).

\bibitem[Ngu05]{OpenSSLDocs}
Jamie Nguyen.
\newblock {OpenSSL certificate authority documentation}, 2005.
\newblock \url{https://tools.ietf.org/html/rfc5054}.

\bibitem[OP17]{Covenants2}
Russell O'Connor and Marta Piekarska.
\newblock Enhancing bitcoin transactions with covenants.
\newblock In {\em Financial Cryptography and Data Security}, pages 191--198.
  Springer International Publishing, 2017.

\bibitem[PR14a]{bip44}
Marek Palatinus and Pavol Rusnak.
\newblock Multi-account hierarchy for deterministic wallets, 2014.
\newblock \url{https://github.com/bitcoin/bips/blob/master/bip-0044.mediawiki}
  (18 May 2020, Last Accessed).

\bibitem[PR14b]{BIP43}
Marek Palatinus and Pavol Rusnak.
\newblock Purpose field for deterministic wallets, 2014.
\newblock \url{https://github.com/bitcoin/bips/blob/master/bip-0043.mediawiki}
  (18 May 2020, Last Accessed).

\bibitem[Rea16]{SecureDataDeletionBook}
Joel Reardon.
\newblock {\em Secure Data Deletion}.
\newblock Springer International Publishing, 2016.

\bibitem[Rub19]{CTVImplementation}
Jeremy Rubin.
\newblock Bitcoin-core checktemplateverify-feb1-workshop branch, 2019.
\newblock
  \url{https://github.com/JeremyRubin/bitcoin/commits/checktemplateverify-feb1-workshop}
  (21 May 2020, Last Accessed).

\bibitem[Rub20]{BIP119}
Jeremy Rubin.
\newblock Checktemplateverify, 2020.
\newblock \url{https://github.com/bitcoin/bips/blob/master/bip-0119.mediawiki}
  (18 May 2020, Last Accessed).

\bibitem[Sch03]{BeyondFear}
Bruce Schneier.
\newblock {\em Beyond Fear: Thinking Sensibly About Security in an Uncertain
  World}.
\newblock Copernicus, 2003.

\bibitem[Sha79]{Shamir1979}
Adi Shamir.
\newblock How to share a secret.
\newblock {\em Communications of the ACM}, volume 22:612–613, 1979.

\bibitem[SHMB20]{Swambo2020cov}
Jacob Swambo, Spencer Hommel, Bob McElrath, and Bryan Bishop.
\newblock Bitcoin covenants: Using pre-signed transactions with secure key
  deletion.
\newblock 2020.
\newblock (Pre-print).

\bibitem[Tim15]{BIP127}
Jorge Timón.
\newblock Simple proof-of-reserves transactions, 2015.
\newblock \url{https://github.com/bitcoin/bips/blob/master/bip-0099.mediawiki}
  (18 May 2020, Last Accessed).

\bibitem[TWMP05]{SRPTLS}
D.~Taylor, T.~Wu, N.~Mavrogiannopoulos, and T.~Perrin.
\newblock {Using the Secure Remote Password (SRP) Protocol for TLS
  Authentication}, 2005.
\newblock \url{https://tools.ietf.org/html/rfc5054}.

\bibitem[Wui12]{BIP32}
Pieter Wuille.
\newblock Hierarchical deterministic wallets, 2012.
\newblock \url{https://github.com/bitcoin/bips/blob/master/bip-0032.mediawiki}
  (18 May 2020, Last Accessed).

\end{thebibliography}

\appendix

\section{Implementation}
\label{Appendix:implementation}

The transaction templates and scripts presented in section \ref{vault-design} have been implemented and tested. The open-source code is available  \href{https://github.com/JSwambo/bitcoin-vault}{here}. The implementation uses the standard functional test framework for bitcoin-core, and explicitly verifies scripts and signatures with the consensus library of bitcoin, libbitcoinconsensus. The tests also demonstrate that the transaction queue (mempool) acceptance policy (which is complex due to transaction replacement features and anti-DoS rules) don't prohibit the proposed unconfirmed transaction dependency chains required for the Vault Custody protocol.  

Additionally, there is a work in progress implementation of the vault wallet software referenced in \ref{sec:system-architecture}. The prototype vault wallet code can be found \href{https://github.com/fmr-llc/Vault-mbed}{here}. The implementation is based off of Stepan Snigirev's `\href{https://diyhpl.us/wiki/transcripts/austin-bitcoin-developers/2019-06-29-hardware-wallets/}{build your own hardware wallet}' workshop. The code is written in C++ and is designed to compile on a standard ST Microelectronics board (model F469NI-DISCO). The board has been used to prototype the functionality for secure key deletion and storage of ACTs. It is also being used to model the vaulting/un-vaulting processes defined in \ref{sec:Vaulting} and \ref{sec:un-vaulting}, although it is not usable as a functional wallet as of yet. The plan is to continue development and release a functional version of the code in the near future. Any and all contributions to bring this forward to a usable wallet state are welcomed.

Finally, an open-source prototype library, \href{https://github.com/kanzure/python-vaults}{python-vaults}, has been developed which provides tools for the construction of pre-signed transaction trees with script templates for vault and push-to-recovery-wallet transactions. The library supports \textit{sharding}, splitting a bitcoin unspent transaction output into multiple new outputs, so that each shard can be un-vaulted according to a specific schedule. Moreover, there is support for OP\_CHECKTEMPLATEVERIFY \cite{BIP119} based covenants and re-vaulting (see appendix \ref{Appendix:revaulting}).

\section{Re-vaulting}
\label{Appendix:revaulting}

\begin{figure}
    \centering
    \includegraphics[width=140mm]{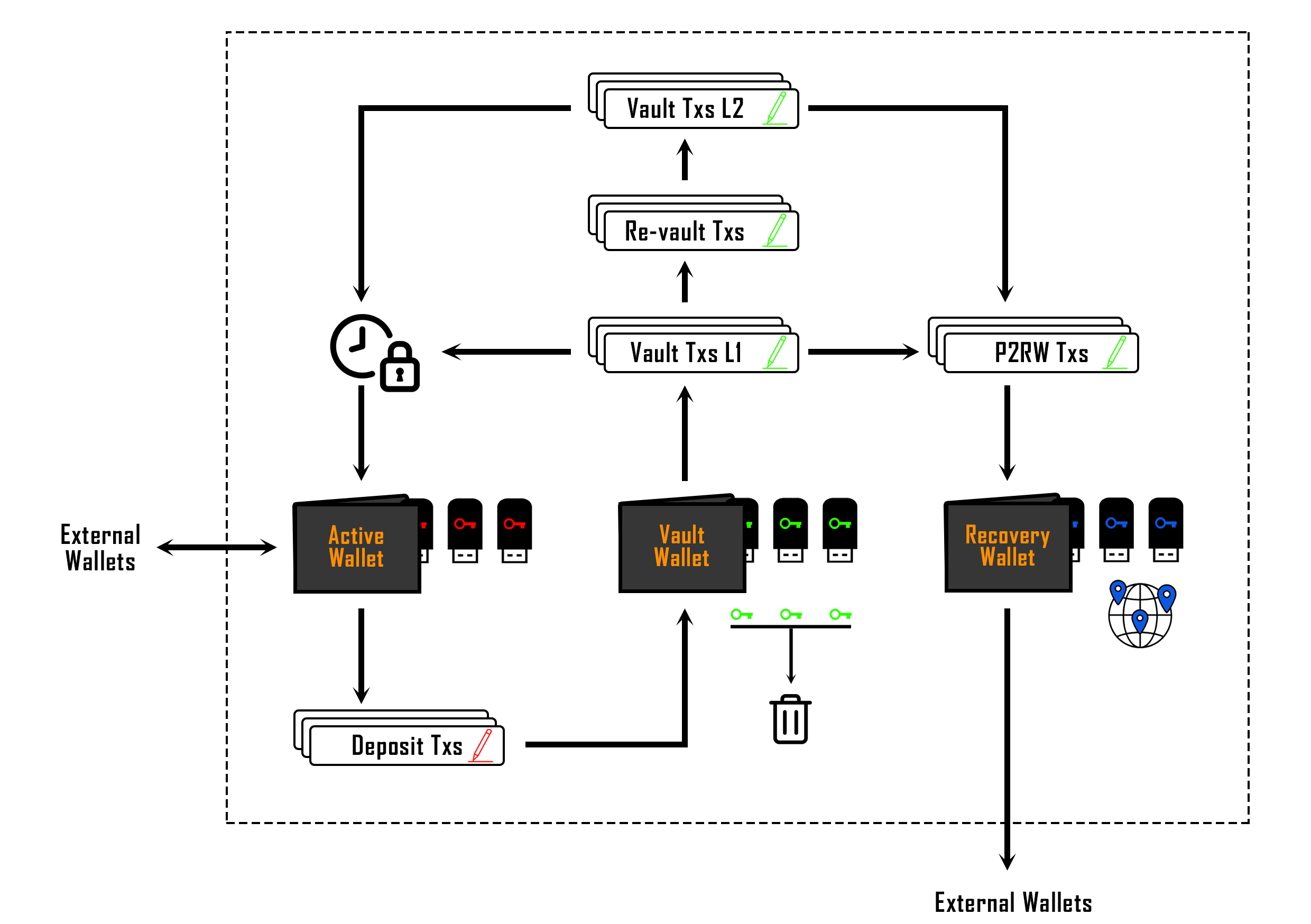} \hspace{2.5cm}
    \caption{Diagram of a modified access control flow diagram that demonstrates the idea of \textit{re-vaulting}, with multiple layers (L1 and L2) of vault transactions. The active wallet can create arbitrary transactions, and is used as the control point for receiving from and spending to wallets that are external to the custody protocol. It also creates deposit transactions to the vault wallet. The vault wallet uses ephemeral keys to sign vault, push-to-recovery-wallet (P2RW), and re-vault covenant transactions. Two layers of vault transactions are shown, where the first layer has three possible spending paths; to a re-vault transaction (where funds are forced to go to the next vault layer), to the active wallet subject to a time-lock, and to a P2RW transaction. The final layer (L2) has two possible spending paths as in the design of section \ref{vault-design}. The recovery wallet is a geographically distributed wallet with high physical security that can sign arbitrary recovery transactions moving funds to an external wallet.}
    \label{fig:Re-vault-AccessControlFlow}
\end{figure}

There has been discussion \cite{moeser2016bitcoin} and development around the idea of \textit{re-vaulting} where each vault transaction has a third possible spending path, one where funds are re-directed to another layer of vault transactions. One may think of this as having an additional recovery option but with devices that are more accessible than the recovery wallet, where the security is provided by a second time-lock enforced by the bitcoin protocol. Multiple vault layers can be used which still enable funds to be pushed to the recovery wallet at any time. A visualisation of the adjusted access control flow is shown in figure \ref{fig:Re-vault-AccessControlFlow}.  It is important to note that a new set of P2RW transactions is needed for each new layer of vault transactions, since the P2RW transactions must reference a specific vault transaction. Moreover, a new covenant type is required that plays a similar role to the P2RW transaction, except that funds are pushed to the next layer of vault transactions instead of the recovery wallet. This is called a \textit{re-vault} transaction.

It is critical that storage for different layers of vault transactions are distinct and separated. This compartmentalization ensures that the compromise of one layer doesn't affect another.  The value of re-vaulting where the new layer of vaults use the same active wallet as the target for un-vaulting is that when storage of one layer of vaults is compromised, funds can be pushed to the next layer. For vault transaction layers that are stored redundantly and without secret sharing, this protects against DoS attacks that would otherwise force the wallet owner to access the recovery wallet.

{}

Alternatively, imagine that each new layer of vault transactions targets a back-up set of HMs that can operate as a new active wallet. The value of this approach is that the wallet owner has an option for a fast recovery, where an attack is thwarted without needing to access the highly secure recovery wallet HMs before continuing normal operation. Of course, this introduces a new set of vulnerabilities since now an attacker can use strategies that compromise the back-up active wallet. This introduces significant complexity to an already complex threat model. 

Dynamic fee allocation in this case doesn't work very well since the chain of committed but unconfirmed covenant transactions is long and it may not be secure to rely on the final transaction to pay for the whole sequence. The simple but inefficient solution is to create each covenant transaction with variants that pay different amounts for fees. The storage costs for this scale poorly with the size of the transaction tree. The costly option is to pay high fees for all transactions and especially high fees for re-vault and P2RW transactions (to better guarantee the protection they offer in the event of compromise).

The overall complexity of the re-vaulting approach is significantly higher, and consideration needs to be made for the storage of each type of covenant transaction at each layer, for fees, for watchtower design (in particular responder watchtowers who would have a new responsibility of deciding whether to re-vault or trigger the recovery process), and for the threat model and risk analysis. Thus a rigorous specification and analysis of re-vaulting is left as a topic for future research. 

\end{document}